\documentclass[twocolumn]{aastex62}
\usepackage[section]{placeins}
\usepackage{longtable}
\usepackage{hyperref}

\graphicspath{{./}{figures/}}

\shorttitle{SN\,2016iet}
\shortauthors{Gomez et al.}

\begin{document}

\title{SN\,2016iet: The Pulsational or Pair Instability Explosion of a Low Metallicity Massive CO Core Embedded in a Dense Hydrogen-Poor Circumstellar Medium}

\correspondingauthor{Sebastian Gomez}
\email{sgomez@cfa.harvard.edu}
	
\author[0000-0001-6395-6702]{Sebastian Gomez}
\affil{Center for Astrophysics \textbar Harvard \& Smithsonian, 60 Garden Street, Cambridge, MA, 02138, USA}

\author[0000-0002-9392-9681]{Edo Berger}
\affil{Center for Astrophysics \textbar Harvard \& Smithsonian, 60 Garden Street, Cambridge, MA, 02138, USA}
	
\author[0000-0002-2555-3192]{Matt Nicholl}
\affil{Institute for Astronomy, University of Edinburgh, Royal Observatory, Blackford Hill, Edinburgh EH9 3HJ, UK}

\author[0000-0003-0526-2248]{Peter K.~Blanchard}
\affil{Center for Astrophysics \textbar Harvard \& Smithsonian, 60 Garden Street, Cambridge, MA, 02138, USA}
	
\author[0000-0002-5814-4061]{V.~Ashley Villar}
\affil{Center for Astrophysics \textbar Harvard \& Smithsonian, 60 Garden Street, Cambridge, MA, 02138, USA}
	
\author{Locke Patton}
\affil{Center for Astrophysics \textbar Harvard \& Smithsonian, 60 Garden Street, Cambridge, MA, 02138, USA}
	
\author{Ryan Chornock}
\affil{Astrophysical Institute, Department of Physics and Astronomy, 251B Clippinger Lab, Ohio University, Athens, OH 45701, USA}

\author[0000-0001-6755-1315]{Joel Leja}
\affil{Center for Astrophysics \textbar Harvard \& Smithsonian, 60 Garden Street, Cambridge, MA, 02138, USA}
	
\author[0000-0002-0832-2974]{Griffin Hosseinzadeh}
\affil{Center for Astrophysics \textbar Harvard \& Smithsonian, 60 Garden Street, Cambridge, MA, 02138, USA}
	
\author[0000-0002-2478-6939]{Philip S.~Cowperthwaite}
\affil{The Observatories of the Carnegie Institution for Science, 813 Santa Barbara Street, Pasadena, CA 91101, USA}

\begin{abstract}
We present optical photometry and spectroscopy of SN\,2016iet (=Gaia16bvd =PS17brq), an unprecedented Type I supernova (SN) at $z=0.0676$ with no obvious analogue in the existing literature. SN\,2016iet exhibits a peculiar light curve, with two roughly equal brightness peaks ($\approx -19$ mag) separated by about 100 days, and a subsequent slow decline by about \mbox{5 mag} in 650 rest-frame days. The spectra are dominated by strong emission lines of calcium and oxygen, with a width of only \mbox{$3400$ km s$^{-1}$}, superposed on a strong blue continuum in the first year. There is no clear evidence for hydrogen or helium associated with the SN at any phase. The nebular spectra exhibit a ratio of \mbox{$L_{\rm [Ca\,II]}/L_{\rm [O\,I]}\approx 4$}, much larger than for core-collapse SNe and Type I SLSNe, but comparable to the so-called Ca-rich transients.  We model the light curves with several potential energy sources: radioactive decay, a central engine, and ejecta-circumstellar medium (CSM) interaction.  Regardless of the model, the inferred progenitor mass near the end of its life (i.e., the CO core mass) is \mbox{$\gtrsim 55$ M$_\odot$} and potentially up to \mbox{$120$ M$_\odot$}, clearly placing the event in the regime of pulsational pair instability supernovae (PPISNe) or pair instability supernovae (PISNe).  The models of CSM interaction provide the most consistent explanation for the light curves and spectra, and require a CSM mass of \mbox{$\approx 35$ M$_\odot$} ejected in the final decade before explosion. We further find that SN\,2016iet is located at an unusually large projected offset ($16.5$ kpc, $4.3$ effective radii) from its low metallicity dwarf host galaxy (\mbox{$Z\approx 0.1$ Z$_\odot$}, \mbox{$L\approx 0.02$ $L_*$}, \mbox{$M\approx 10^{8.5}$ M$_\odot$}), supporting the interpretation of a PPISN/PISN explosion.  In our final spectrum at a phase of about 770 rest-frame days we detect weak and narrow H$\alpha$ emission at the location of the SN, corresponding to a star formation rate of \mbox{$\approx 3\times 10^{-4}$ M$_\odot$ yr$^{-1}$}, which is likely due to a dim underlying galaxy host or an \ion{H}{2} region. Despite the overall consistency of the SN and its unusual environment with PPISNe and PISNe, we find that the inferred properties of SN\,2016iet challenge existing models of such events.
\end{abstract}


\section{Introduction} \label{sec:intro}

Over the past decade, wide-field optical surveys have been successful at uncovering new types of rare transient events, such as superluminous supernovae (SLSNe; e.g., \citealt{quimby11, chomiuk11}), tidal disruption events (e.g., \citealt{gezari09, chornock14, arcavi14}), fast blue optical transients (e.g., \citealt{drout14}), and calcium-rich transients (e.g., \citealt{kasliwal12}). At the same time, some predicted types of SNe remain elusive, in particular pair-instability SNe (PISNe; \citealt{barkat67}) and pulsational pair-instability SNe (PPISNe; \citealt{woosley07}).

PISNe are the predicted explosions of massive stars with zero age main sequence (ZAMS) masses of \mbox{$130-260$ M$_\odot$}. They are caused by the high core temperatures in massive stars, which lead to the production of electron-positron pairs that in turn lead to contraction of the core, followed by explosive oxygen burning that eventually unbinds the star, leaving no remnant behind \citep{barkat67, rakavy67,kasen11}. The luminous SN\,2007bi was claimed to be a PISN based on its slowly declining light curve and claimed large mass of radioactive $^{56}$Ni \citep{galyam09}.  However, \citet{nicholl13} suggested that SN\,2007bi was powered by a magnetar central engine with an ejecta mass of only \mbox{$7$ M$_\odot$}. The spectra, metallicity, and observational features of SN\,2007bi has been shown to be inconsistent with a PISN \citep{jerkstrand16,dessart13, mazzali19}. To date, there have been no definitive cases of PISN explosions.

For stars with slightly lower ZAMS masses of \mbox{$95-130$ M$_\odot$}, the progenitor can experience multiple non-destructive pair instability episodes that expel material prior to the final core collapse. These pulses can lead to shell collisions that power a SN-like transient -- a PPISN -- while the inner layers contract and temporarily resume stable burning \citep{woosley07}. The ejecta produced in the ultimate core collapse explosion (if successful) can also interact with shells ejected \mbox{$\sim 10^4$ years} prior, powering a luminous and long lasting SN \citep{woosley07,aguilera18}. There are several observed SNe with unusual photometric or spectroscopic features that have been interpreted as signs of PPISNe. The hydrogen-poor SN\,2010mb \citep{benami14} exhibited a long lasting light curve, narrow oxygen emission lines, and a persistent blue continuum.  The hydrogen-rich iPTF14hls had a light curve with at least 5 distinct peaks over multiple years, absorption lines of constant velocity, and a sub-solar metallicity host \citep{arcavi17}. The hydrogen-poor SLSN iPTF16eh showed evidence for a light echo from a shell whose properties might match some PPISN models \citep{lunnan18}. 

Against this backdrop, we present detailed optical observations of SN\,2016iet, a transient initially discovered by the Gaia Photometric Science Alerts Team, and also recovered by the Catalina Real-Time Transient Survey and the Pan-STARRS Survey for Transients. Our follow-up observations over the past 3 years place the event at a redshift of $z=0.0676$ and reveal highly unusual light curves, spectra, and environment. The light curve has a double-peaked structure of roughly equal brightness peaks of \mbox{$\approx -19$ mag} separated by 100 days, followed by a long duration and slow decline, and a total radiated energy of a \mbox{${\rm few}\times 10^{50}$ erg}. The spectra are dominated by intermediate velocity (\mbox{$\approx 3400$ km s$^{-1}$}) lines of calcium and oxygen superposed on an initially blue continuum, and show no clear evidence for hydrogen and helium associated with the SN. Finally, we find that SN\,2016iet is located at an unusually large offset of $16.5$~kpc ($4.3$ effective radii) from its low metallicity dwarf host galaxy. Weak and narrow H$\alpha$ emission at the SN position points to a potential underlying dwarf galaxy or a single \ion{H}{2} region.  The combination of these unusual properties prompts us to explore a wide range of models for the energy source powering SN\,2016iet: radioactive decay, a central engine (magnetar or fallback accretion), and circumstellar interaction.  We find that regardless of the model, the inferred progenitor mass (CO core) is \mbox{$\gtrsim 55$ M$_\odot$} (and perhaps as high as \mbox{$\approx 120$ M$_\odot$}), placing this event in the realm of PPISNe and PISNe. 

The structure of the paper is as follows. In \S\ref{sec:observations} we present the discovery of SN\,2016iet and our follow-up observations. In \S\ref{sec:analysis} we describe the multi-color and bolometric light curves, as well as the spectra. In \S\ref{sec:modeling} we present our modeling of the light curve, and in \S\ref{sec:host} we describe the host galaxy. In \S\ref{sec:discussion} and \S\ref{sec:conclusion} we discuss possible interpretations of this event and present our key conclusions. Throughout the paper we assume a flat $\Lambda$CDM cosmology with \mbox{$H_{0} = 69.3$ km s$^{-1}$ Mpc$^{-1}$}, $\Omega_{m} = 0.286$, and $\Omega_{\Lambda} = 0.712$ \citep{hinshaw13}.

\section{Discovery and Observations} \label{sec:observations}

\subsection{Discovery}
\begin{figure*}
	\begin{center}
		\includegraphics[width=0.9\textwidth]{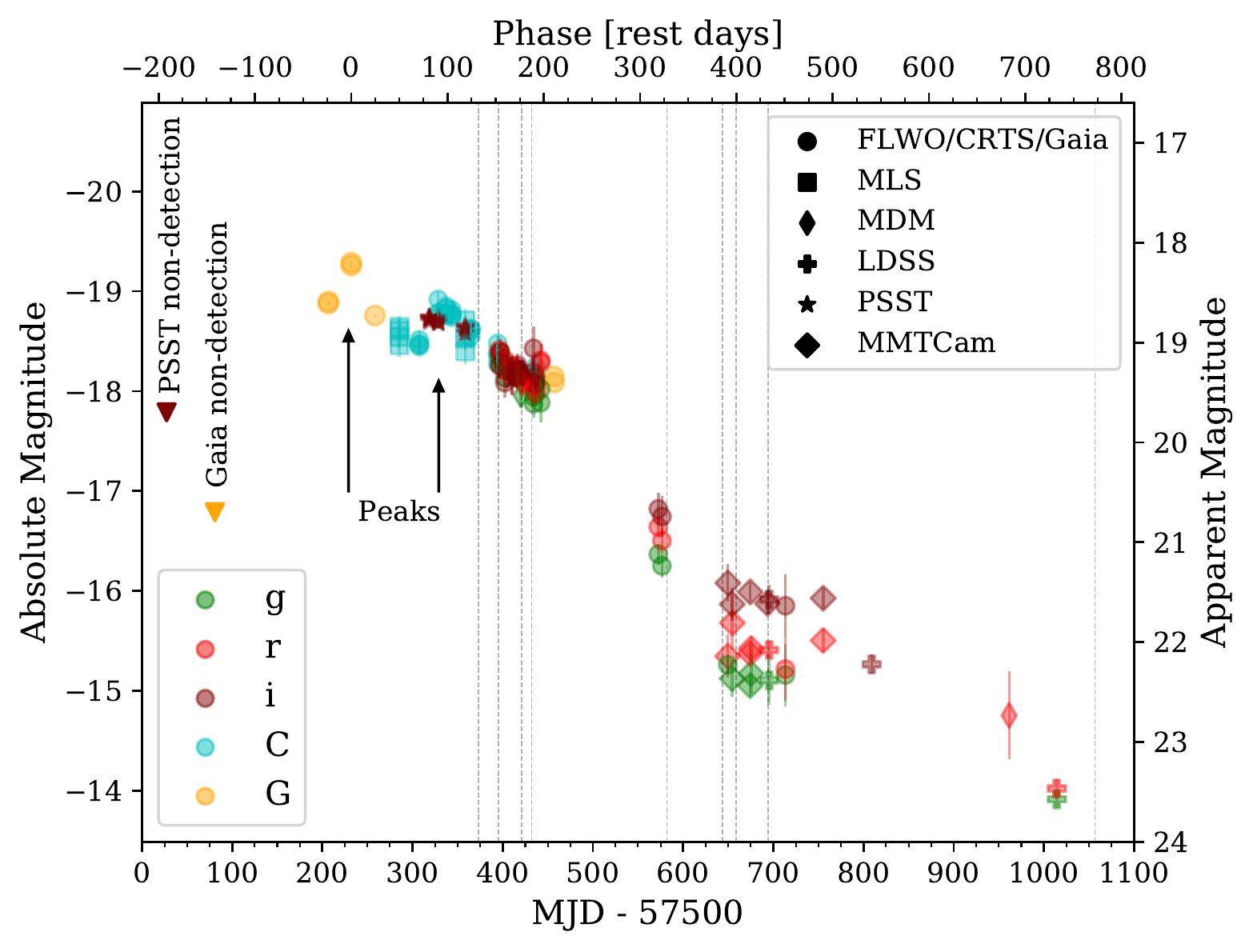}
		\caption{Optical light curve of SN\,2016iet in the $gri$, Gaia-\textit{G}, and CRTS unfiltered-C bands. The photometry is corrected for Galactic extinction. The vertical dashed lines indicate the epochs of spectroscopic observations.  The two distinct peaks are marked by arrows. 
			\label{fig:lightcurve}}	
	\end{center}
\end{figure*}

SN\,2016iet was first discovered by the Gaia Science Alerts (GSA; \citealt{wyrzykowski12}) on 2016 November 14 (${\rm MJD}=57706.596$) and announced two days later with the designation Gaia16bvd with a magnitude of $G=18.61$ at  coordinates R.A.=${\rm 12^h32^m33^s.23}$, decl.=$+27^\circ07'15''.49$ (J2000). The transient was subsequently recovered by the Catalina Real-Time Transient Survey (CRTS; \citealt{drake09}) on 2017 January 5 with a magnitude of $C=19.03$ and was given two separate designations (MLS170316:123233+270716 and CSS170201:123233+270716), and by the Pan-STARRS Survey for Transients (PSST; \citealt{huber15}) on 2017 March 6 with a magnitude of $i=18.72$, and the designation PS17brq.  The GSA and PSST provide upper limits from images prior to discovery of $G>20.7$ on 2016 July 11 and $i>19.7$ on 2016 May 19; see Figure~\ref{fig:lightcurve}.  We find a $3\sigma$ upper limit of $r\gtrsim 23.0$ mag through forced photometry at the position of SN\,2016iet in archival Pan-STARRS1 $3\pi$ images.

As part of our search for exotic transients, we flagged SN\,2016iet as a host-less transient given the PS1/$3\pi$ non-detection at the SN position.  We eventually found that it resides at a large separation of $\approx 11''$ from a galaxy with a half-light radius of $\approx 2.6''$ that we identify as its host based on a common redshift (\S\ref{sec:host}). This prompted us to obtain an optical spectrum on 2017 April 30 with the Inamori-Magellan Areal Camera and Spectrograph (IMACS; \citealt{dressler11}) on the Magellan Baade 6.5-m Telescope. The spectrum revealed strong calcium and oxygen features, weak Si features, and no evidence for H or He lines, which in principle lead to a Type~Ic designation.  However, SN\,2016iet does not resemble any typical Type~Ic SNe, and we therefore adopt a general classification of a Type~I SN. 

To determine the redshift of SN\,2016iet we measure the wavelengths of the strongest emission lines in two individual spectra: \mbox{[\ion{Ca}{2}] $\lambda\lambda7291,7324$}, \mbox{\ion{O}{2} $\lambda7320,7330$}, and \mbox{\ion{Ca}{2} $\lambda\lambda\lambda8498,8542,8662$}. We simultaneously fit a Gaussian profile for each emission line in each spectrum and derive $z = 0.0676\pm 0.0002$.

\begin{figure}[t]
	\begin{center}
		\includegraphics[width=\columnwidth]{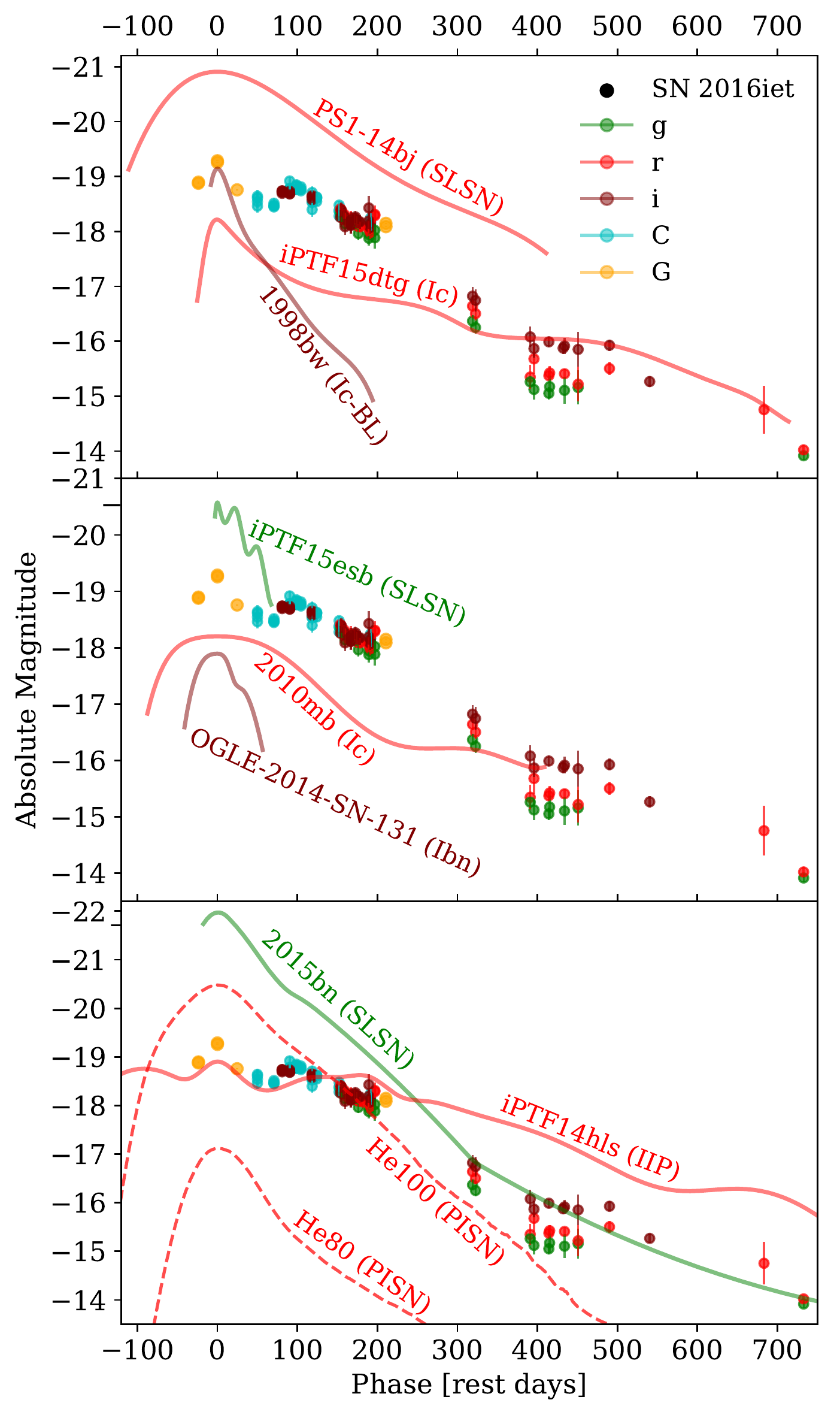}
		\caption{The light curve of SN\,2016iet in comparison to other types of hydrogen-poor and/or helium-poor SNe: SN\,2010mb \citep{benami14}, iPTF15esb \citep{yan17}, iPTF15dtg \citep{taddia18}, SN\,2015bn \citep{nicholl16a,nicholl16b,nicholl18c}, PS1-14bj \citep{lunnan16}, SN\,1998bw \citep{galama98}, and OGLE-2014-SN-131 \citep{karamehmetoglu17}. We also include a comparison to the He80 and He100 PISN models from \cite{kasen11}, and to the hydrogen-rich PPISN candidate iPTF14hls \citep{arcavi17}. The comparison light curves have been smoothed for clarity, and shifted in phase to match the first peak of SN\,2016iet. \label{fig:comparisons}}
	\end{center}
\end{figure}

\subsection{Optical Photometry}

We obtained images of SN\,2016iet in the $gri$ filters from several instruments: KeplerCam on the 1.2-m telescope at Fred Lawrence Whipple Observatory (FLWO), Templeton on the 1.3-m McGraw-Hill Telescope at MDM Observatory, the Low Dispersion Survey Spectrograph (LDSS3c; \citealt{stevenson16}) on the Magellan Clay 6.5-m telescope at Las Campanas Observatory, and MMTCam on the MMT 6.5-m telescope. We processed the images using standard IRAF\footnote{\label{IRAF}IRAF is written and supported by the National Optical Astronomy Observatories, operated by the Association of Universities for Research in Astronomy, Inc. under cooperative agreement with the National Science Foundation.} routines, and performed photometry with the {\tt daophot} package.  Instrumental magnitudes were measured by modeling the PSF from each image using reference stars and subtracting the model from the target. For calibration, we estimated individual zero-points by measuring the magnitudes of field stars and comparing to photometric AB magnitudes from the PS1/$3\pi$ catalog. The uncertainties reported here are the combination of the photometric uncertainty and the uncertainty in the zero-point determination.  The resulting photometry is summarized in Table~A\ref{tab:photometry}. SN\,2016iet was also observed on several epochs by Gaia, PSST, and CRTS in the \textit{G}, $i$, and unfiltered C bands, respectively (these data are included in Table~A\ref{tab:photometry}). 

The full light curve of SN\,2016iet, spanning to January 2019, is shown in Figure~\ref{fig:lightcurve}, where we define phase 0 to be the date of the brightest magnitude in \textit{G} band, \mbox{${\rm MJD} = 57732$}.  We use this reference point throughout the paper, with all days quoted in the rest frame unless otherwise specified. All of the photometry has been corrected for Galactic extinction with \mbox{$A_V = 0.041 \pm 0.001$ mag} \citep{SF2011}. We determine that host subtraction and host extinction corrections are not necessary, based on measurements of the host extinction being consistent with zero and the large separation of SN\,2016iet from its host (\S\ref{sec:host}). All the photometry collected for this work is made available on the Open Supernova Catalog \footnote{\label{ref:osc}\url{https://sne.space/}} \citep{guillochon17}.

\subsection{Optical Spectroscopy}

We obtained nine epochs of low-resolution optical spectroscopy at phases of 132 to 772 days. We used the IMACS and LDSS3c spectrographs on the Magellan telescopes, the Blue Channel \citep{schmidt89} and Binospec \citep{fabricant03} spectrographs on the MMT, the Gemini-North Multi-Object Spectrograph (GMOS; \citealt{hook04}), and the Ohio State Multi-Object Spectrograph (OSMOS; \citealt{martini11}) on the 2.4-m Hiltner Telescope at MDM Observatory. All the spectra were obtained with the slit oriented along the parallactic angle. The details of the spectroscopic observations are provided in Table~A\ref{tab:spectroscopy}.

We processed the spectra using standard IRAF routines with the {\tt twodspec} package. The spectra were bias and flat corrected, the sky background was modeled and subtracted from each image, and the one-dimensional spectra were optimally extracted, weighing by the inverse variance of the data. A wavelength calibration was applied using a HeNeAr lamp spectrum taken near the time of each science image. Relative flux calibration was applied to each spectrum using a standard star taken on the same night. The spectra were then calibrated to absolute flux using our FLWO $gri$ photometry as a reference, where a normalization constant was applied to each spectrum to match the expected flux from the photometry using the {\tt PYPHOT} Python package. Lastly, the spectra were corrected for Galactic extinction and transformed to the rest frame of SN\,2016iet. All the spectroscopy collected for this work is made available on the Open Supernova Catalog $^{\ref{ref:osc}}$ \citep{guillochon17}.

The spectra obtained on 2017 November 25 and 2018 January 21 with the Blue Channel Spectrograph also include the host galaxy. The one-dimensional spectra of the host were extracted in the same way as outlined above and are discussed in \S\ref{sec:host}.

\clearpage
\section{Observed Properties of the Light Curves and Spectra}\label{sec:analysis} 

\begin{figure}
	\begin{center}
		\includegraphics[width=\columnwidth]{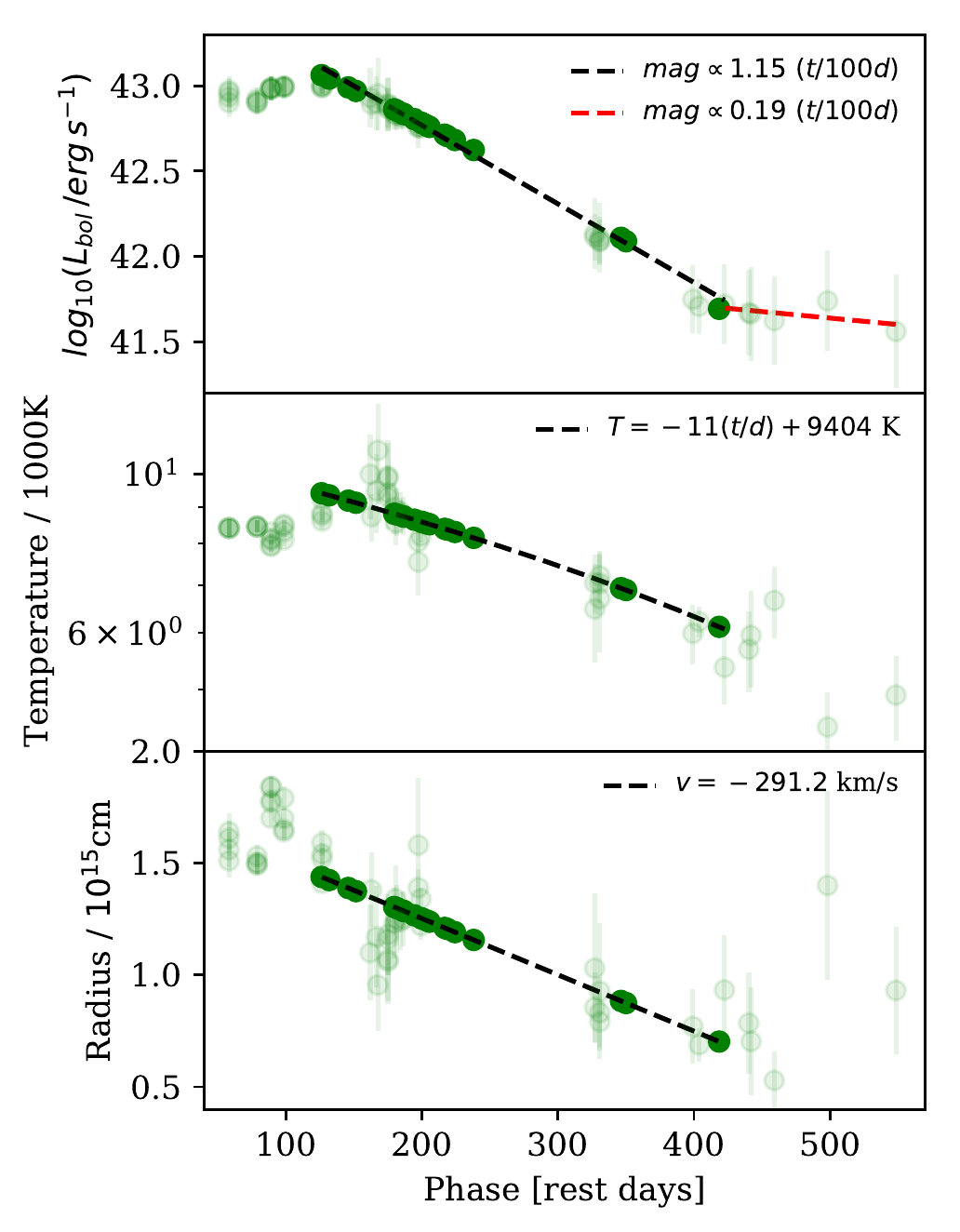}
		\caption{Bolometric light curve ({\it Top}), blackbody temperature ({\it Middle}), and photospheric radius ({\it Bottom}) of SN\,2016iet. In all panels the solid green points show the result of a simultaneous fit to all epochs with multi-band data assuming a linear decline in temperature and radius.  The dashed lines indicate the rate of decline and the time range over which we fit the data. The light green points are the result of interpolating the photometry in all bands to a common grid.  We do not include the first peak due to the lack of color information. \label{fig:bolometric}}
	\end{center}
\end{figure}

\begin{figure*}
	\begin{center}
		\includegraphics[width=0.9\textwidth]{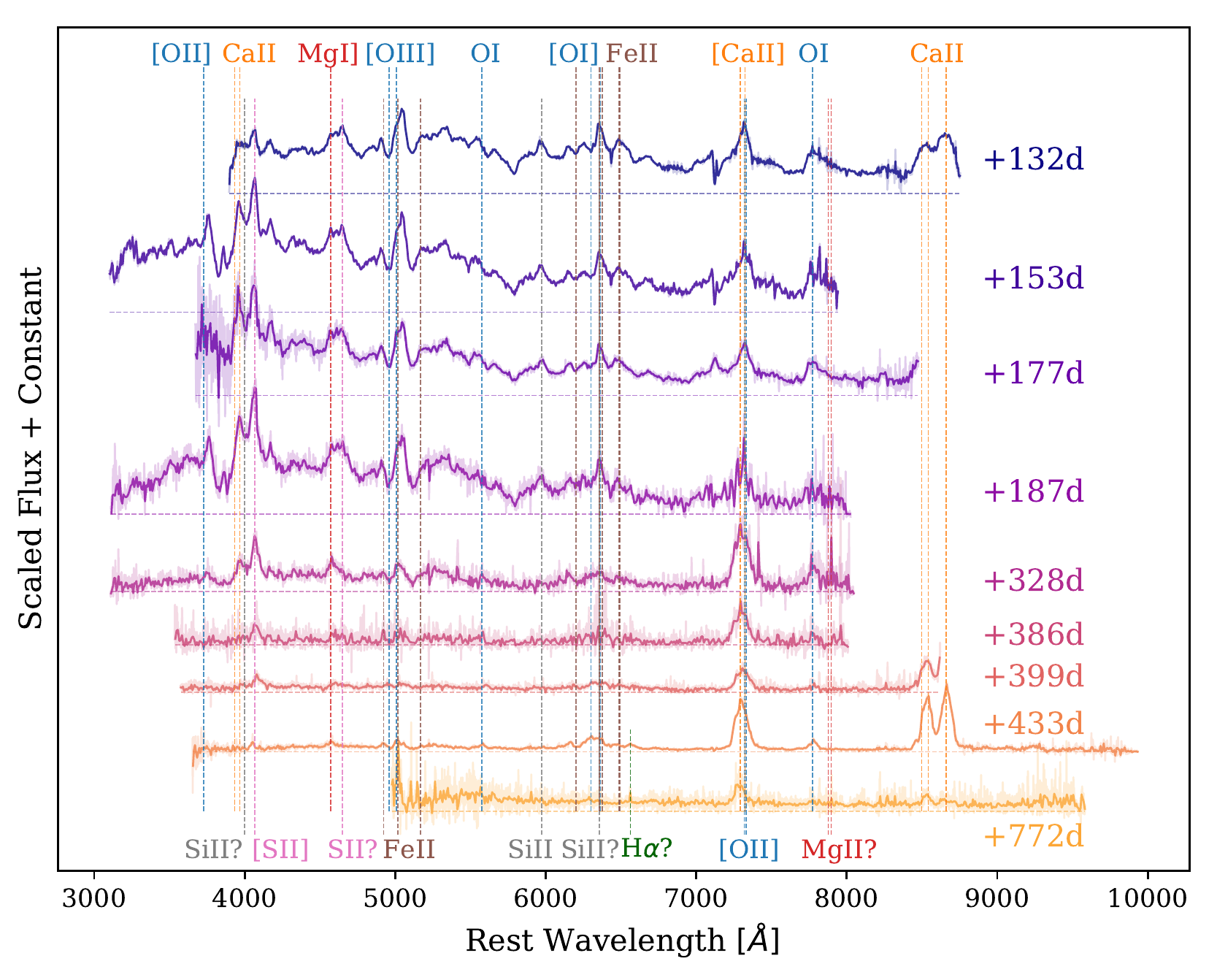}
		\caption{Optical spectra of SN\,2016iet spanning a rest-frame phase of 132 to 772 days relative to the first peak. The spectra are corrected for Galactic extinction and shifted to the rest-frame of SN\,2016iet using $z=0.0676$. In some epochs we binned the spectra for clarity; the unbinned spectra are shown as shaded regions. The dashed horizontal lines indicate the zero level of each spectrum.  The dominant lines in the spectra are marked. \label{fig:spectra}}
	\end{center}
\end{figure*}

\subsection{Multi-band Light Curve}

The multi-band light curves of SN\,2016iet are shown in Figure~\ref{fig:lightcurve}. The most striking features are the two distinct peaks, and the overall long duration of the event. The first peak, detected by Gaia, appears to have a roughly symmetrical shape, rising by 0.5 mag in $\approx 30$ days to a peak of $M_G\approx -19.2$, and declining to $M_G\approx -18.7$ after 25 days, when the SN was detected by the CRTS and PSST. These data reveal a second peak, with \mbox{$M_C\approx -18.7$ mag}, about 100 days after the first peak.

Following the second peak, the bolometric light curve declines at a rate of $\approx 0.011$ mag day$^{-1}$ for about 300 days and then flatten for about 100 days before resuming a declining behavior.  In addition to the overall fading we also find that the emission becomes redder with time, from initial colors soon after the second peak of $g-r\approx r-i\approx 0.1$ mag, to \mbox{$g-r\approx 0.35$} and \mbox{$r-i\approx 0.8$ mag} one year after the second peak.  As we show below, this color evolution is primarily driven by an initially blue continuum, which eventually fades, leaving behind mainly strong nebular emission lines of calcium and oxygen. Our last data points have a magnitude $\approx 23.5$ mag, deeper by about 0.5 mag than the pre-existing PS1/$3\pi$ upper limits at the SN position.  Since the SN is still detectable from the ground 2.5 years after discovery, future ground and space observations will reveal the continued light curve behavior, and potentially any underlying faint host.

In Figure~\ref{fig:comparisons} we compare the light curves of SN\,2016iet to those of both normal and unusually long-lived Type~I SNe.  SN\,2016iet exhibits a distinct evolution from any of these events. Even its first peak alone is wider than the broad-lined Type~Ic SN\,1998bw \citep{galama98} or even the longest duration Type~Ibn SN detected to date, OGLE-2014-SN-131 \citep{karamehmetoglu17}. Other events match a similar overall timescale (e.g., the SLSNe SN\,2015bn and PS1-14bj; \citealt{lunnan16,nicholl16a, nicholl16b, nicholl18c}), but are either much less luminous or more rapidly fading from the initial brighter peak.  The closest light curve analogues are the Type~Ic events iPTF15dtg \citep{taddia18} and SN\,2010mb \citep{benami14}, but even those do not share the double-peaked structure of SN\,2016iet.  The SLSNe LSQ14bdq \citep{nicholl15} and iPTF15esb \citep{yan17} exhibit undulations fin their light curves, but those are much less pronounced and occur on a much shorter timescales ($\sim 20$ days) than the double peaks in SN\,2016iet. The Type~IIP SN iPTF14hls \citep{arcavi17} has a light curve of longer duration than SN\,2016iet and it shows undulations on timescales comparable to SN\,2016iet, but it is a hydrogen-rich event. To conclude, we do not find a clear analogue to SN\,2016iet in the existing SN sample with comparable behavior, brightness, and duration.

In Figure~\ref{fig:comparisons} we also include a comparison to the PISN light curve models for 80 and 100 M$_\odot$ bare helium cores from \cite{kasen11}. We find the peak brightness of SN\,2016iet is bounded by these two models, but the models lack the double-peaked structure and fade much more rapidly.

\subsection{Bolometric Light Curve} 
\label{sec:bolometric}

We calculate the bolometric light curve of SN\,2016iet, its photospheric radius, and temperature evolution by modeling the spectral energy distribution (SED) in each epoch by fitting a blackbody function using the {\tt Superbol} code \citep{nicholl18b}. We further extrapolate the blackbody SED to account for the missing coverage in the UV and NIR. At phases earlier than 75 days we assume a constant color to account for missing multi-band photometry. The resulting bolometric light curve, photospheric radius, and temperature evolution are shown in Figure~\ref{fig:bolometric}.

For the second light curve peak we find a peak bolometric luminosity of \mbox{$\approx 1.2\times 10^{43}$ erg s$^{-1}$}. The integrated radiated energy over the 650 rest-frame days following the second peak is $E_{\rm rad}\approx 2\times 10^{50}$ erg.  This does not include the first peak of the light curve for which we have no color information.

We have $gri$-band photometry for phases between about 125 and 420 days, where we infer a steadily decreasing photospheric radius and temperature. Fitting all of these data simultaneously with an assumed linear decline, we find that at the time of the second peak the photospheric temperature was \mbox{$T=9400\pm 500$ K}, and the radius was \mbox{$R=(1.4\pm 0.1)\times 10^{15}$ cm}, with their respective linear decline fits shown in Figure~\ref{fig:bolometric}. The bolometric light curve follows a decline of \mbox{$0.0115\pm 0.0002$ mag day$^{-1}$} until day 420, and subsequently flattens out to a decline rate of \mbox{$0.0019\pm 0.0018$ mag day$^{-1}$} at a phase of $420-580$ days.

The data during the first peak are limited to only the Gaia $G$-band. To assess its bolometric luminosity and radiated energy we explore two extreme cases. First, assuming a constant temperature of about \mbox{8400 K} during the first peak (the earliest temperature measurement at 84 days), we find a photospheric radius of \mbox{$\approx 1.8\times 10^{15}$ cm} and a peak luminosity of \mbox{$\approx 1.2\times10^{43}$ erg s$^{-1}$}. Second, assuming a radius equal to the smallest photospheric radius (\mbox{$\approx 7\times 10^{14}$ cm} at \mbox{420 days}) we find a temperature of \mbox{$\approx 24000$ K} and a peak bolometric luminosity of \mbox{$\approx 1.3\times 10^{44}$ erg s$^{-1}$}. This later luminosity corresponds to a total radiated energy during the first peak of \mbox{$E_{\rm rad}\sim 10^{50}$ erg}, for an estimated rise time of 30 days.

\begin{figure}
	\begin{center}
		\includegraphics[width=\columnwidth]{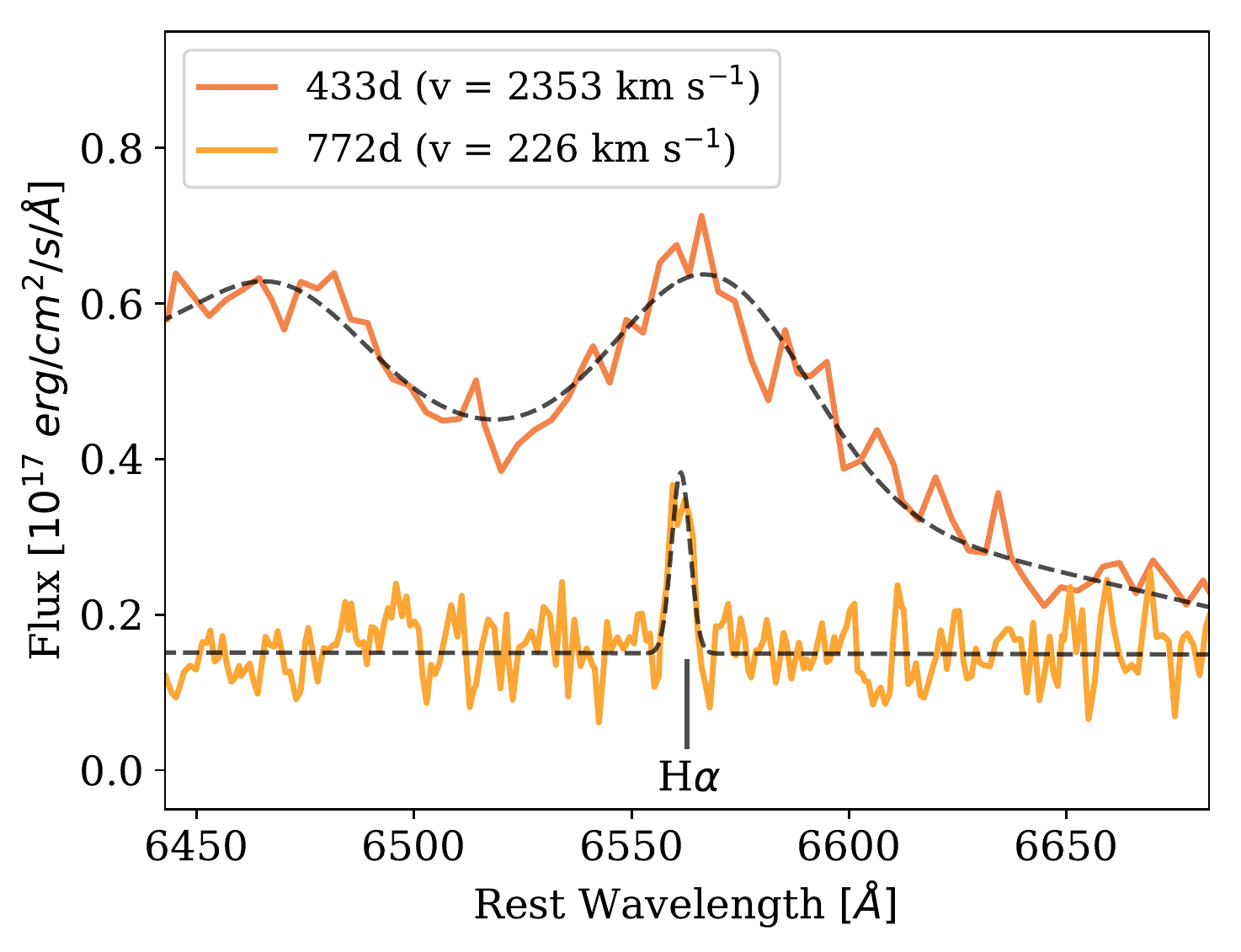}
		\caption{Zoom-in on the location of H$\alpha$ in the last two spectra of SN\,2016iet. The spectrum at a phase of 433 days shows a broad line at the location of H$\alpha$. We fit this feature, and surrounding region, with a double Gaussian component plus a flat continuum and find a velocity width of \mbox{$2350$ km s$^{-1}$}. At a phase of 722 days we detect narrow unresolved emission at the location of H$\alpha$ likely due to star formation at the location of the supernova. We fit this line with a single Gaussian and find a velocity width \mbox{$\lesssim 230$ km s$^{-1}$}. \label{fig:halpha}}
	\end{center}
\end{figure}

\begin{figure}
	\begin{center}
		\includegraphics[width=\columnwidth]{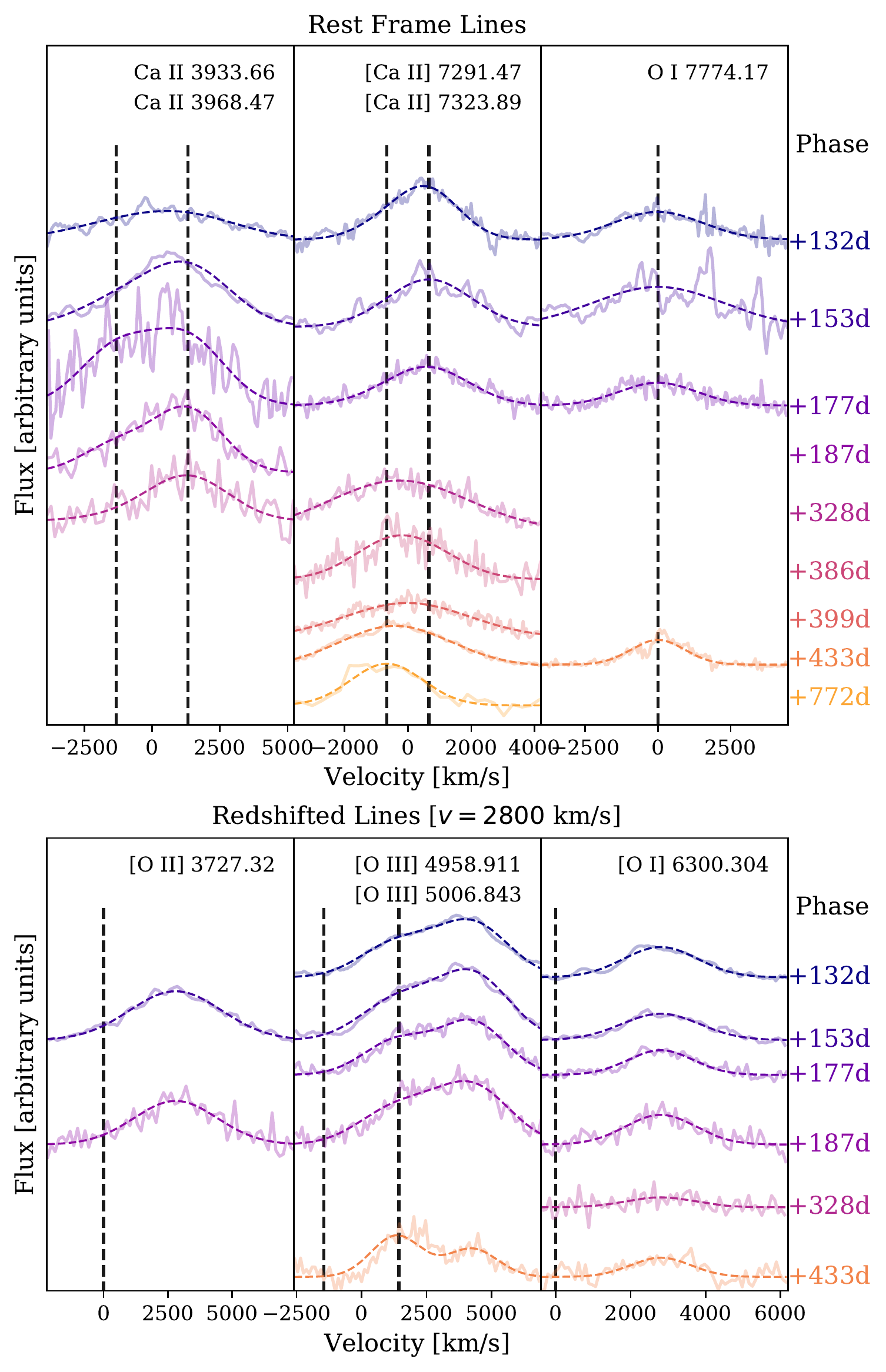}
		\caption{Zoom-in on the most prominent emission lines in the spectra of SN\,2016iet. In some cases we exclude phases that lack spectral coverage or have a poor signal-to-noise ratio at the location of specific lines. {\it Top}: Emission lines of \ion{Ca}{2}, [\ion{Ca}{2}], and \ion{O}{1}, which define the rest-frame of the SN. {\it Bottom}: Forbidden oxygen lines ([\ion{O}{1}], [\ion{O}{2}], and [\ion{O}{3}]), which are observed to be systematically redshifted by \mbox{$\approx 2800$ km s$^{-1}$} with respect to the SN rest-frame. \label{fig:line_profiles}}
	\end{center}
\end{figure}

\subsection{Spectral Features} 
\label{sec:spectra}

Our spectra of SN\,2016iet span 132 to 772 rest-frame days and are shown in Figure~\ref{fig:spectra}. The  spectra earlier than about 330 days are dominated by intermediate width emission lines of \ion{Ca}{2}, \ion{O}{1}, [\ion{Ca}{2}], [\ion{O}{1}], [\ion{O}{2}], [\ion{O}{3}], \ion{Mg}{1}], and a series of blended Fe lines near $5000$\AA. A strong blue continuum is also present.  Beyond about 380 days, the blue continuum fades, although it does not completely disappear, and the spectra are dominated by strong emission lines of calcium, and oxygen, and weaker lines of sulfur and magnesium, leading to the color evolution seen in the light curves. 

\begin{figure}
	\begin{center}
		\includegraphics[width=\columnwidth]{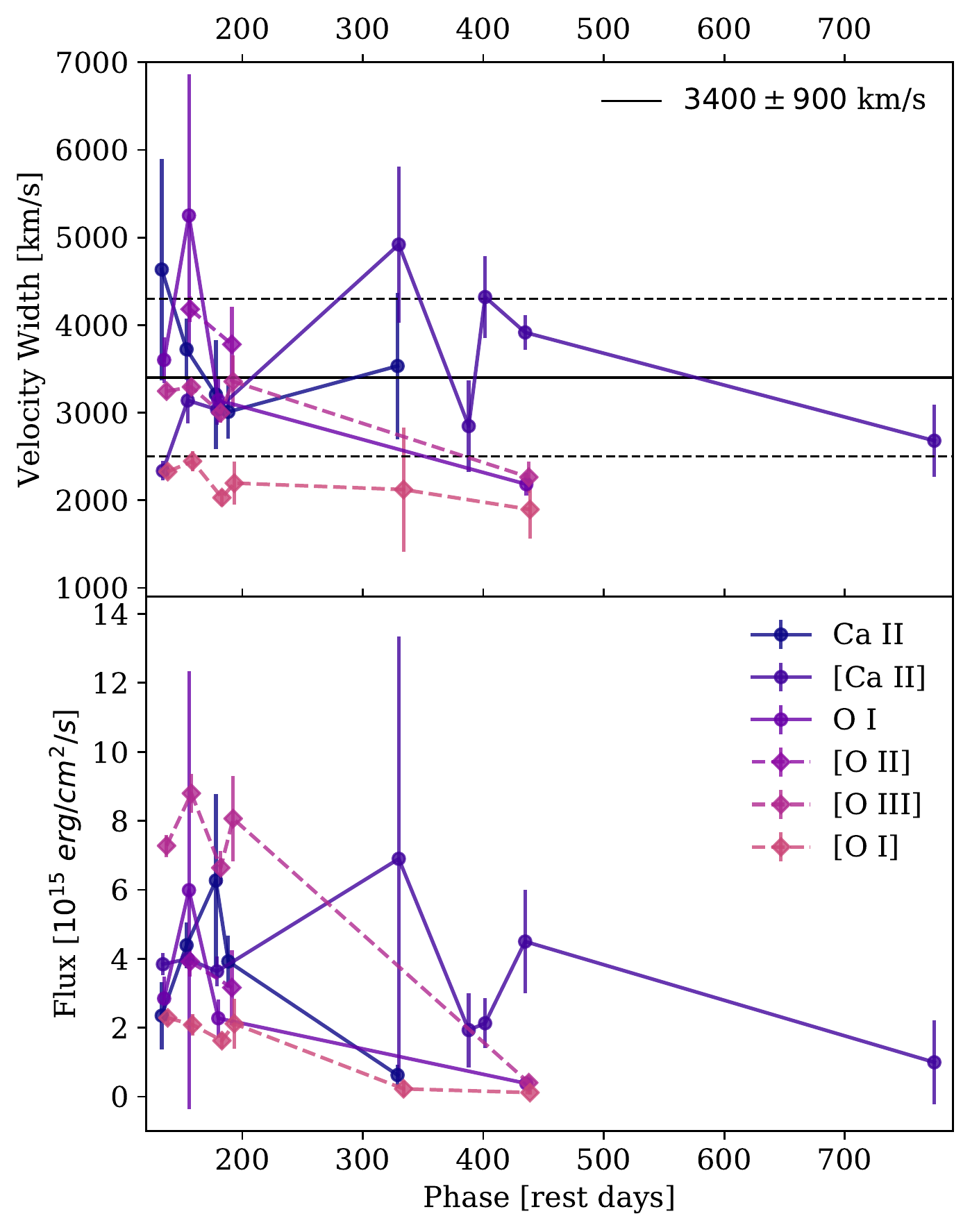}
		\caption{{\it Top}: Velocity width of the most prominent calcium and oxygen lines. No obvious evolution in the line width is observed over a period of nearly 2 years. The solid and dashed horizontal lines mark the mean and standard deviation of the  velocity width, $3400\pm 900$ km s$^{-1}$. {\it Bottom}: Fluxes of the same calcium and oxygen lines. The dashed color lines indicate the forbidden oxygen lines, which are redshifted by $\approx 2800$ km s$^{-1}$ from the rest of the lines. The velocities and fluxes are calculated from Gaussian fits to the line profiles shown in Figure~\ref{fig:line_profiles}. \label{fig:line_fluxes}}
	\end{center}
\end{figure}

The spectra show no clear evidence for hydrogen or helium associated with the SN, although we detect a feature at the location of H$\alpha$ ($5\sigma$ confidence level) exclusively in the spectrum at 433 days. The origin of this feature is uncertain due to its faintness and presence in only a single spectrum. It could be due to ejecta interaction with a thin layer of hydrogen, or not be associated to H$\alpha$ at all, but instead be produced by either  \ion{Fe}{2} or \ion{N}{2}.

In our final spectrum at a phase of 772 days we instead detect a narrow line at the location of H$\alpha$, shown in Figure~\ref{fig:halpha}. This line is unresolved ($v\lesssim 230$ km s$^{-1}$), and most likely associated with underlying star formation at the location of the SN. We measure its luminosity to be \mbox{$(3.9\pm 1.5)\times 10^{37}$ erg s$^{-1}$}, which corresponds to a star formation rate of \mbox{${\rm SFR}\approx 3\times 10^{-4}$ M$_\odot$ yr$^{-1}$} \citep{kennicutt98}. This value is close to the lower limit for dwarf galaxies, but similar to \ion{H}{2} regions in the SMC \citep{kennicutt98,hony15}. In the preceding spectrum (phase of 433 days) the SN was a few times brighter in the continuum, explaining the absence of this line in earlier spectra.

The \ion{O}{1} $\lambda$7774 line is present in all the spectra and exhibits a broad red shoulder that is most pronounced in the early epochs, potentially due to contamination from \ion{Mg}{2} $\lambda\lambda$7877,7896 (e.g., \citealt{nicholl18a}). The emission lines in the complex feature around $4000$ \AA\ are likely due to \ion{Ca}{2} $\lambda\lambda3934,3968$. We see an unusual absorption doublet at $\sim 3850$ \AA\ where the lines have a velocity width of $3600\pm 550$ km s$^{-1}$, this feature could be caused by a P-Cygni profile of the \ion{Ca}{2} doublet, but the exact nature of the line remains uncertain.

To measure the velocity width and integrated flux of the various lines we fit single or double Gaussian profiles to single and doublet lines, respectively; see Figure~\ref{fig:line_profiles}.  We use the following lines:  [\ion{Ca}{2}] $\lambda\lambda7291,7324$, \ion{Ca}{2} $\lambda\lambda3934,3968$, \ion{O}{1} $\lambda7774$, [\ion{O}{1}] $\lambda6300$, [\ion{O}{2}] $\lambda3727$, and [\ion{O}{3}] $\lambda\lambda4959,5007$, and find an average velocity of $3400\pm 900$ km s$^{-1}$, with no evidence for significant change with time (Figure~\ref{fig:line_fluxes}). The forbidden oxygen lines are all redshifted by $\approx 2800$ km s$^{-1}$ relative to the permitted oxygen lines and both the permitted and forbidden calcium lines (bottom panel of Figure~\ref{fig:line_profiles}). We present a possible interpretation for this shift in \S\ref{sec:redshifted_lines}.

Beginning from our first spectrum at a phase of 132 days, the lines of \ion{Ca}{2} and \ion{O}{1} increase in flux by a factor of about $2-3$ during the first 30 days, and then decline for the next 200 days by a factor of $\sim 10$. The [\ion{O}{2}] and [\ion{O}{3}] line fluxes remain constant for the first 50 days, and then [\ion{O}{3}] declines by a factor of $\sim 15$ for the next 250 days.  The [\ion{O}{1}] line flux appears to decrease monotonically by a factor of $\sim 10$ for 200 days, starting with our first observation, and then remains constant for 100 days. The strongest line is \ion{O}{1} for most of our observations, until day 400, when it is overtaken by the [\ion{Ca}{2}]$\lambda\lambda7291,7324$ doublet. The flux of this [\ion{Ca}{2}] line remains constant within a factor of 2 throughout, until the last spectrum at phase 772 days.

We measure the flux ratio of the \mbox{[\ion{Ca}{2}] $\lambda\lambda7291,7324$} to \mbox{[\ion{O}{1}] $\lambda\lambda6300,6364$} doublets at a phase of 433 days to be \mbox{$L_{\rm [Ca\,II]}/L_{\rm [O\,I]} = 3.8\pm 1.0$}. This is higher than for Type~IIb/Ib/Ic SNe, which have typical ratios of $\lesssim 2$ \citep{milisavljevic17}. This high ratio is also uncommon for Type~I SLSNe, although comparable values have been measured for LSQ14an and PTF10hgi at late times \citep{nicholl18a}.  However, in those SLSNe the \mbox{[\ion{Ca}{2}] $\lambda\lambda7291,7324$} doublet was argued to be actually dominated by \mbox{[\ion{O}{2}] $\lambda\lambda7319,7330$} emission \citep{jerkstrand17}; we do not believe this to be the case for SN\,2016iet since we would expect this [\ion{O}{2}] doublet to be redshifted like all of the other forbidden oxygen lines (Figure~\ref{fig:line_profiles}).

The large ratio of \mbox{$L_{\rm [Ca\,II]}/L_{\rm [O\,I]}$} is instead reminiscent of the calcium-rich gap transients \citep{kasliwal12}, which have values of \mbox{$\approx 2.5-11$} \citep{lunnan17,milisavljevic17}. However, these transients are significantly less luminous, and evolve on much faster timescales, reaching peak magnitudes of only about $ -16$ within $\sim 15$ days after explosion, and entering the nebular phase in \mbox{$\lesssim 100$ days}.  Incidentally, calcium-rich gap transients have large projected offsets from their host galaxies \citep{lunnan17}, much like SN\,2016iet (\S\ref{sec:host}), but unlike SN\,2016iet they are typically found in massive galaxies \citep{de2018, lunnan17}.

\begin{figure}[t!]
	\begin{center}
		\includegraphics[width=\columnwidth]{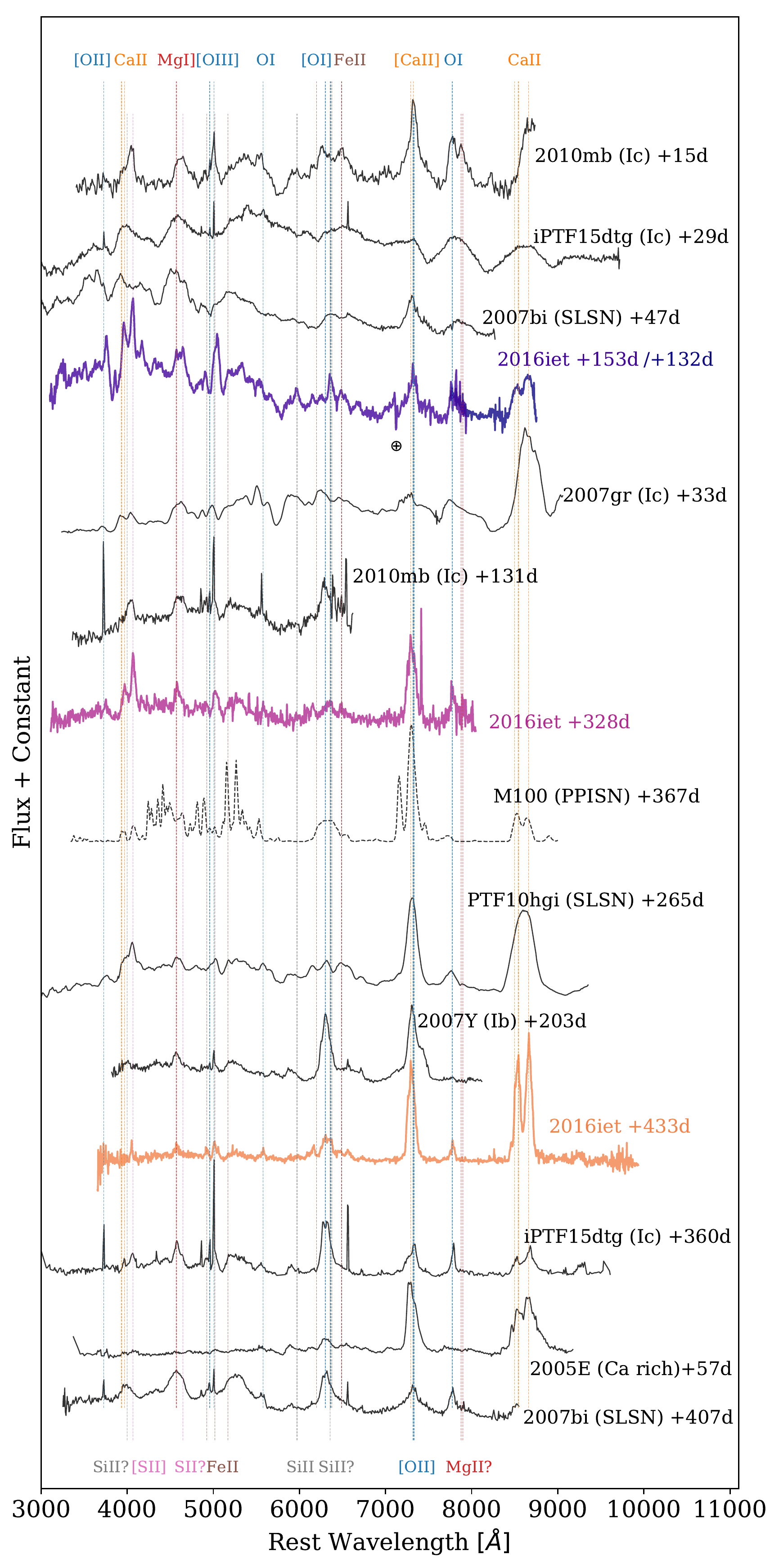}
		\caption{Comparison of three representative spectra of SN\,2016iet (colored lines), including the early photospheric phase with a strong blue continuum (purple), the transition to the nebular phase at 328 days when the continuum begins to fade (pink), and the late nebular phase at 433 days with strong calcium and oxygen emission lines (yellow). We compare the evolution of SN\,2016iet to other hydrogen- and/or helium-poor SNe (black). Strong telluric lines are marked with $\oplus$. The vertical lines mark the locations of dominant lines in the spectra of SN\,2016iet. \label{fig:spectra_comparison}}
	\end{center}
\end{figure}

\subsection{Spectral Comparisons} 
\label{sec:spectra_compare}

In Figure~\ref{fig:spectra_comparison} we compare the spectra of SN\,2016iet to those of various Type~I SNe. The earliest spectra of SN\,2016iet, at a phase of \mbox{$130-150$ days} exhibit emission lines of mainly calcium and oxygen, similar to the Type~Ic SN\,2007gr at a phase of 33 days \mbox{\citep{valenti08}}, although the emission lines in SN\,2007gr are broader, with velocities of \mbox{$\sim 5000$ km s$^{-1}$}. SN\,2007gr also does not exhibit the strong blue continuum present in SN\,2016iet. SN\,2016iet shows a much slower spectral evolution than typical Type~Ic SNe, reaching the nebular phase only at about 400 days. The spectrum of SN\,2016iet at a phase of 350 days is similar to the 200-day spectrum of SN\,2007Y, a Type~Ib SN with high velocity \ion{Ca}{2} lines that was suggested to be powered by CSM interaction \citep{stritzinger09}. The spectrum of SN\,2016iet at 462 days also resembles that of calcium-rich gap transients, for example SN\,2005E \citep{perets10,lunnan17} but at a phase of only about 60 days, further emphasizing the fast evolution of calcium-rich gap transients as compared to SN\,2016iet. We also show a comparison to PTF10hgi \citep{inserra13,quimby18}, a SLSN with one of the highest calcium to oxygen ratios (although as indicated above this ratio might be misinterpreted due to a possibly dominant [\ion{O}{2}] line).

The Type~IIP SN iPTF14hls has a light curve with similar peak brightness and undulations to SN\,2016iet, yet their spectra are completely different, where iPTF14hls is dominated by strong hydrogen lines \citep{arcavi17}. The Type~Ic SN iPTF15dtg also has a light curve similar in duration to SN\,2016iet, but a spectrum unlike that of SN\,2016iet that is instead similar to typical Type~Ic SNe \citep{taddia18}.

We show SN\,2007bi as an example of a slowly-evolving SLSN \citep{galyam09, nicholl13}. The nebular spectra of SN\,2007bi exhibit strong features of Fe-group elements, which we do not observe in SN\,2016iet. In SN\,2016iet we observe much narrower lines than in SN\,2007bi, closer to the velocities of \mbox{$\sim 2000 - 4000$ km s$^{-1}$} expected from a PISN \citep{woosley17,mazzali19}. We also plot a PISN model of a 100 M$_\odot$ He core from \citet{mazzali19}. However, the model differs significantly from the spectra of SN\,2016iet, mainly due to a series of narrow iron lines.

SN\,2016iet is comparable only to SN\,2010mb, with the caveat that SN\,2010mb shows narrow emission lines of \mbox{[\ion{O}{3}] $\lambda\lambda4959,5007$}, \mbox{[\ion{O}{2}] $\lambda3727$}, and \mbox{[\ion{O}{1}] $\lambda5577$} with a velocity with of \mbox{$\sim 800$ km s$^{-1}$} \citep{benami14}. SN\,2010mb also showed emission lines of calcium, oxygen, and magnesium with similar velocities to SN\,2016iet, and a blue continuum at early phases, although not as prominent as in SN\,2016iet \citep{benami14}.  Unfortunately, the last spectrum of SN\,2010mb was at only 131 days after explosion, preventing a direct comparison to the nebular spectra of SN\,2016iet.

\subsection{Redshifted Oxygen Lines:  A Light Echo?}
\label{sec:redshifted_lines}

We find that all of the forbidden oxygen lines are redshifted by $\approx 2800$ km s$^{-1}$ relative to the permitted oxygen lines and the permitted and forbidden calcium lines (Figure~\ref{fig:line_profiles}).  This redshift is comparable to the velocity width of the lines. The redshifted lines are also symmetric around the redshifted velocity, rather than exhibiting a skewed velocity profile. The critical densities necessary to observe forbidden calcium and oxygen lines vary only by a factor of $\sim 2$, with $n = 3.8\times10^{6}$ cm$^{-3}$ for \mbox{[\ion{O}{1}] $\lambda\lambda6300,6364$}, and $n = 8.8\times10^{6}$ cm$^{-3}$ for \mbox{[\ion{Ca}{2}] $\lambda\lambda7291,7324$} \citep{jerkstrand17}.  This mild difference is unlikely to account for the different kinematics of the forbidden oxygen and calcium lines.

Instead it appears that the permitted and forbidden oxygen lines must be arising from physically distinct regions, with the latter showing a lower density and an apparent recession. One possibility is that the forbidden lines are in fact an echo of the SN light from a circumstellar shell not yet reached by the ejecta, expanding at the observed redshift of $2800$ km s$^{-1}$; such a velocity could arise in a giant eruption such as from the pulsational pair instability. If this shell was ejected before the SN, that would account for its lower density.  Recently, \cite{lunnan18} detected an echo in the \ion{Mg}{2}$\lambda$2800 resonance line from a CSM shell around the SLSN iPTF16eh. In that case, the line showed a temporal evolution from blueshifted to redshifted, due to the difference in light travel time to the observer between the front and back of the shell. In the case of SN2016iet, the lines are at all times redshifted, indicating that if they arise from an echo, we have missed the blueshifted phase. This sets constraints on the distance to the shell: to already see photons reflected from the far side of the shell in our earliest spectrum at 132 days, the inner radius of the shell must be located $\lesssim 1.7\times10^{17}$ cm from the star. For a velocity of $2800$ km s$^{-1}$, this implies an ejection $\lesssim 20$ years before explosion.

\begin{deluxetable}{ll}
	\tablecaption{MOSFiT Parameter Definitions \label{tab:parameters}}
	\tablehead{\colhead{Parameter} & \colhead{Definition}}
	\startdata
	$M_{\text{ej}}$      & Ejecta mass  \\
	$M_{\text{Ni}}$      & Radioactive nickel mass  \\
	$f_{\text{Ni}}$      & Nickel mass as fraction of the  ejecta mass  \\
	$L_1$      & Engine power for fallback model at 1 sec   \\
	$t_{\text{on}}$      & Fallack power input constant up to this time \\
	$v_{\text{ej}}$      & Ejecta velocity  \\
	$E_{k}$              & Ejecta kinetic energy   \\
	$M_{\text{CSM}}$     & CSM mass      \\
	$R_{0}$              & CSM inner radius  \\
	$\rho$               & CSM density  \\
	$M_{\text{NS}}$      & Neutron star mass   \\
	$P_{\text{spin}}$    & Magnetar Spin   \\
	$B_{\perp}$          & Magnetar magnetic field strength \\
	$t_{\text{exp}}$     & Explosion time relative to first data point  \\
	$T_{\text{min}}$     & Photosphere temperature floor  \\
	$n_{H,\text{host}}$  & Column density in the host galaxy \\
	$A_{V, \text{host}}$ & Extinction in the host galaxy   \\
	$\kappa$             & Optical opacity \\
	$\kappa_{\gamma}$    & Gamma-ray opacity  \\
	$\sigma$             & Uncertainty required for $\chi^2_r=1$ \\
	\enddata
\end{deluxetable}

More difficult for this scenario is that the line remains visible over 300 days, which for a sharp injection of luminosity would require a very thick shell with a width ($\Delta R$) several times greater than its inner radius. However, this may instead be a function of the fact that the light curve of SN\,2016iet remains relatively flat for $\sim 200$ days. Given the long duration of the SN, it is then surprising that we do not see a blueshifted component. The unexplained absorption feature at $3850$ \AA\ may offer a clue -— if this is from blueshifted \mbox{Ca II H \& K} lines, the implied velocity of the absorbing material is $\sim 5000$ km/s, distinct from the redshifted forbidden oxygen lines. These lines could be coming from a distinct shell, or in fact not be related to the \mbox{Ca II H \& K} lines. We defer further discussion of these lines until future detailed modeling, which is beyond the scope of this paper.

In summary, SN\,2016iet shows a significant blue continuum, constant intermediate width emission lines of calcium and oxygen with velocities of \mbox{$3400\pm 900$ km s$^{-1}$}, a high ratio of calcium to oxygen, no signs of hydrogen or helium, a peculiar set of redshifted forbidden oxygen lines that might be the result of a light echo, and a slowly evolving spectrum dissimilar to any normal Type~Ic SN.

\section{Light Curve Modeling} 
\label{sec:modeling} 

Due to the unusual photometric and spectrosopic nature of SN\,2016iet, we explore a wide range of physical models to explain its origin, including radioactive decay, a magnetar and fallback accretion central engine, and circumstellar interaction. We fit the multi-band light curves using the Modular Open-Source Fitter for Transients ({\tt MOSFiT}) Python package, a flexible MCMC code designed to model the light curves of transients using a variety of different power sources \citep{guillochon18}. We run each MCMC using an implementation of the {\tt emcee} sampler \citep{foreman13}. We test for convergence by ensuring that the models reach a Potential Scale Reduction Factor of $<1.2$ \citep{gelman92}, which corresponds to about $20,000-30,000$ steps with 200 walkers, depending on the specific model.  The parameters of the various models are summarized in Table~\ref{tab:parameters}.

\begin{deluxetable*}{ccccc}
	\tablecaption{Best-fit Parameters for the Radioactive Decay Models \label{tab:radioactive}}
	\tablehead{\colhead{Parameter} & \colhead{Prior}  & \colhead{Full Light Curve} & \colhead{First Peak}  & \colhead{Units}}
	\startdata
	$M_{\text{ej}}                $ & $[1, 260]$         & $ 64^{+27}_{-20}  $ & $ 5.7^{+12.6}_{-2.3}   $ & M$_\odot$              \\
	$f_{\text{Ni}}                $ & $[0.001, 1]$     & $ 0.06 \pm 0.02         $ & $ 0.24^{+0.2}_{-0.1}  $ &                        \\
	$M_{\text{Ni}}^{\dagger}      $ &                 & $ 3.7^{+0.5}_{-0.3}       $ & $ 1.5 ^{+0.5}_{-0.2}         $ & M$_\odot$              \\
	$V_{\text{ej}}                $ & $3400 \pm 900^a$  & $ 3474^{+845}_{-820}    $ & $ 19675^{+4084}_{-8593} $ & km s$^{-1}$            \\
	$E_{k}^{\dagger}              $ &                 & $ 4.4^{+4.8}_{-2.5}     $ & $ 13.7^{+33}_{-10}        $ & $10^{51}$ erg s$^{-1}$ \\
	$t_{\text{exp}}               $ & $[0, 500]$         & $ 118^{+18}_{-23}       $ & $ 17.9^{+8.0}_{-2.8}    $ & days                   \\
	$T_{\text{min}}               $ & $[1000, 15000]$  & $ 7638^{+405}_{-382}   $ & $ 3758^{+5071}_{-2181}  $ &   K                      \\
	$\log{(n_{H,\text{host}})}    $ & $[16, 23]$         & $ 17.0 \pm 0.07          $ & $ 17.9 \pm 1.7          $ & cm$^{-2}$              \\
	$A_{V, \text{host}}^{\dagger} $ &                 & $ < 0.002  $ & $< 0.04  $ & mag                    \\
	$\kappa                       $ & $[0.1, 0.2]$       & $ 0.15 \pm 0.03         $ & $ 0.14 \pm 0.03         $ &  cm$^2$g$^{-1}$          \\
	$\log{(\kappa_{\gamma})}      $ & $[-4, 4]$          & $ 1.6^{+1.6}_{-1.7}   $ & $  1.5\pm 1.9          $ & cm$^2$g$^{-1}$         \\
	$\log{\sigma}                 $ & $[-4, 2]$          & $ -0.53 \pm 0.03        $ & $ -2.8\pm 1.0 $ &                        \\
	\enddata
	\tablecomments{Best model parameters, prior ranges, and 1$\sigma$ error bars for the realizations shown in Figure~\ref{fig:radioactive}. See Table~\ref{tab:parameters} for parameter definitions.}
	\tablenotetext{\dagger}{These parameters were not fit for, but were calculated using all the posterior distribution samples of the fitted parameters.}
	\tablenotetext{a}{Prior only used for Full Light Curve model, the prior for the First Peak model was set to $1000-25000$ km s$^{-1}$.}
\end{deluxetable*}

\subsection{Radioactive Decay}
\label{sec:radioactive}

We begin with a radioactive nickel-cobalt decay model, which is the dominant power source in normal Type~I SNe, although we note that this model cannot strictly account for the double-peaked structure of SN\,2016iet.  The input luminosity in this model is described by:
\begin{equation}
L_\gamma = M_{\rm Ni}\left(\epsilon_{\rm Ni}e^{-t/\tau_{\rm Ni}}+\epsilon_{\rm Co}e^{-t/\tau_{\rm Co}}\right)\ ,
\end{equation}
where $t$ is time, $M_{\rm Ni}$ is the initial mass of $^{56}$Ni, \mbox{$\epsilon_{\rm Ni} = 6.45\times10^{43}$ erg s$^{-1}$ M$_\odot^{-1}$} and $\epsilon_{\rm Co}=1.45\times10^{43}$ erg s$^{-1}$ M$_\odot^{-1}$ are heating rates for $^{56}$Ni and $^{56}$Co, respectively, and $\tau_{\rm Ni}=8.8$ days and $\tau_{\rm Co}=111.3$ days are the half-lives of $^{56}$Ni and $^{56}$Co, respectively \citep{nadyozhin94}.

\begin{figure}[t!]
	\begin{center}
		\includegraphics[width=\columnwidth]{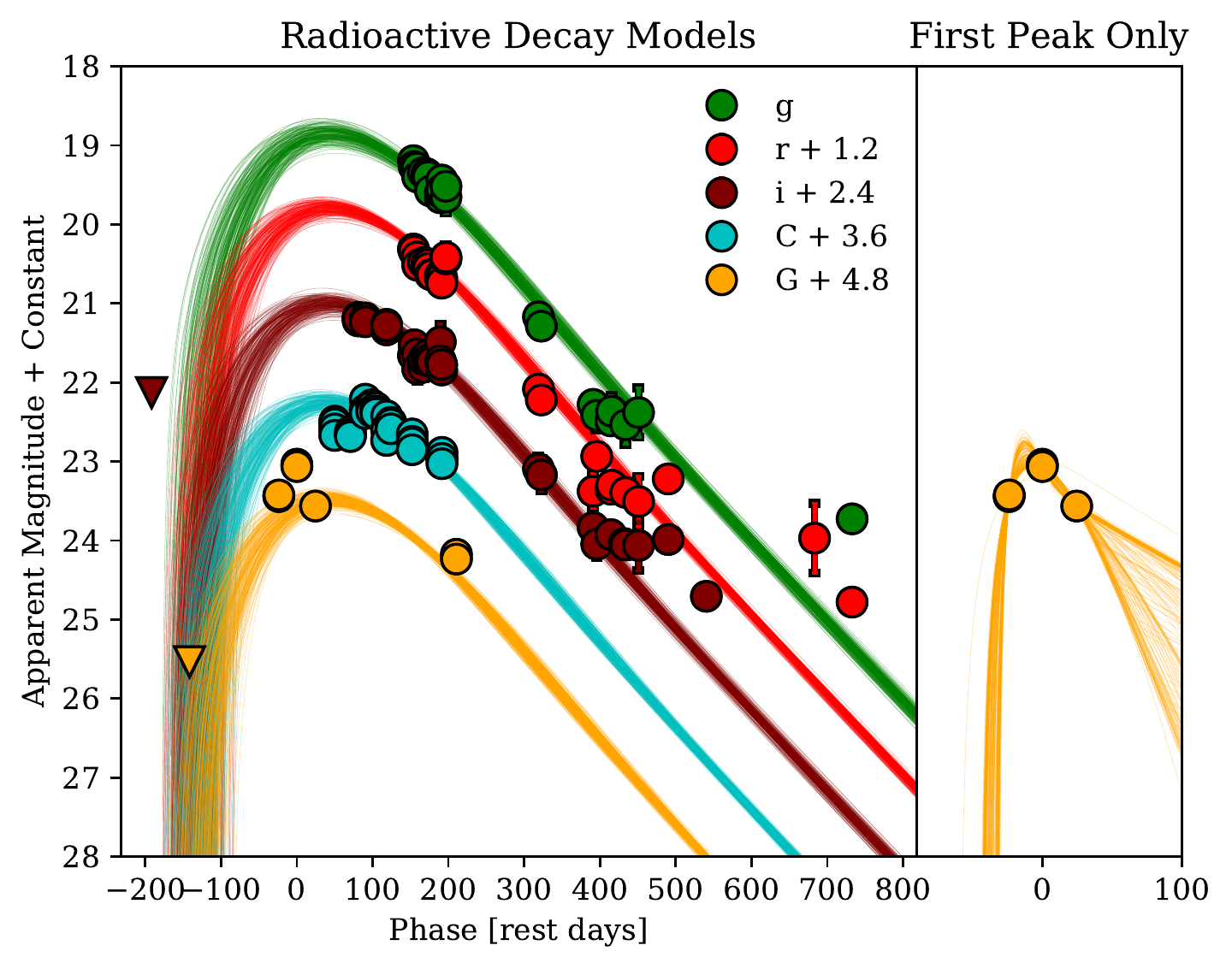}
		\caption{Light curve of SN\,2016iet with radioactive decay models for the full light curve ({\it Left}) and only the first peak ({\it Right}). Each line is a sample realization of the most likely models generated from {\tt MOSFiT}. The resulting parameters for both models are provided in Table~\ref{tab:radioactive}, and the models are discussed in \S\ref{sec:radioactive}. \label{fig:radioactive}}
	\end{center}
\end{figure}

\begin{figure}[t!]
	\begin{center}
		\includegraphics[width=\columnwidth]{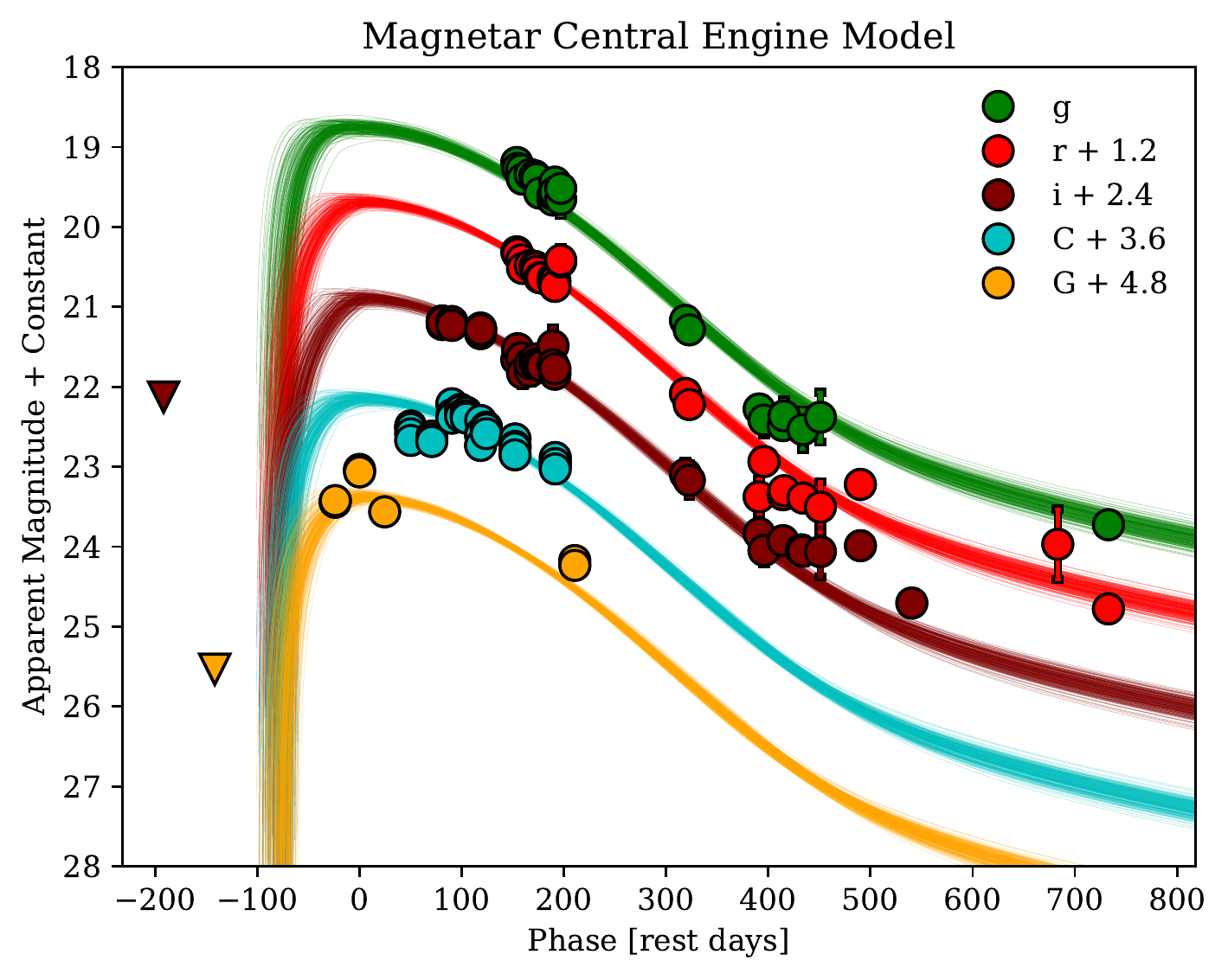}
		\includegraphics[width=\columnwidth]{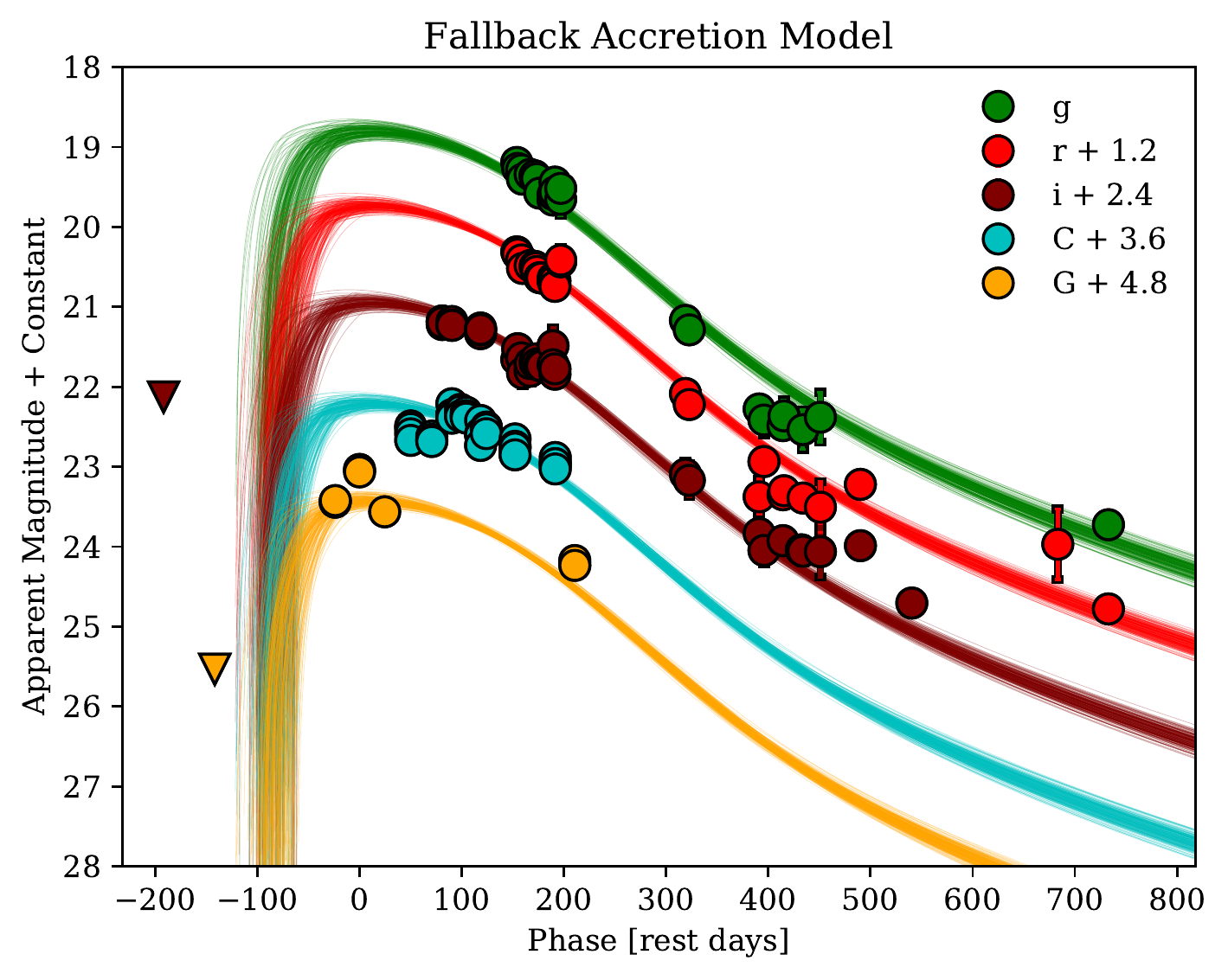}
		\caption{Light curve of SN\,2016iet with a magnetar ({\it Top}) and fallback accretion ({\it Bottom}) central engine model.  Each line is a sample realization of the most likely models generated from {\tt MOSFiT}. The resulting parameters are provided in Table~\ref{tab:magnetar}, and the models are discussed in \S\ref{sec:magnetar} and \S\ref{sec:fallback}. \label{fig:magnetar}}
	\end{center}
\end{figure}

The most likely models are shown in Figure~\ref{fig:radioactive}. The models exhibit a broad light curve around the time of the first observed peak. But as noted above, the model cannot account for the double-peaked structure, or for the flattening beyond 420 days. By the time of our latest data points the model underpredicts the observed brightness by about 1.5 mag.

For this model we find \mbox{$M_{\rm ej}\approx 64$ M$_\odot$}, \mbox{$M_{\rm Ni}\approx 3.7$ M$_\odot$}, \mbox{$v_{\rm ej}\approx 3500$ km s$^{-1}$} (where the prior of the velocity was set by the observed width of the emission lines), and \mbox{$E_K\approx 4.4\times 10^{51}$ erg}; the full list of parameters and their associated uncertainties is provided in Table~\ref{tab:radioactive}.  The model has a nickel fraction of $f_{\rm Ni}=0.06\pm 0.02$, consistent with broad-lined Type~Ic SNe, which have typical values of \mbox{$f_{\rm Ni}\approx 0.05-0.15$} \citep{taddia19}. However, this contrasts with the lack of evidence for a large fraction of Fe-group elements in the spectra of SN\,2016iet, and the presence of a strong blue continuum.  We therefore consider radioactive decay an unlikely model for the full light curve of SN\,2016iet.

Since we have no spectroscopic or color constraints during the first peak, and its duration is more comparable to normal Type~I SNe (although still broader than even SN\,1998bw, Figure~\ref{fig:comparisons}), we explore a model in which only this portion of the data is explained by radioactive decay. In this case, we do not impose a limiting prior on $v_{\rm ej}$ due to the lack of spectroscopic data. The best fit is shown in the right panel of Figure~\ref{fig:radioactive}. We find \mbox{$M_{\rm ej} \approx 6$ M$_\odot$}, \mbox{$M_{\rm Ni}\approx 1.5$ M$_\odot$} (\mbox{$f_{\rm Ni}\approx 0.24$}), \mbox{$v_{\rm ej}\approx 2.0\times 10^4$ km s$^{-1}$}, and \mbox{$E_K\approx 1.4\times 10^{52}$ erg s$^{-1}$} (Table~\ref{tab:radioactive}). In this case both $f_{\rm Ni}$ and $E_K$ are much larger than for Type~Ic SNe \citep{taddia19}.  We therefore consider this model unlikely, although it cannot be strictly ruled out. Naturally, this model still requires a distinct energy source for the remainder of the light curve.

\subsection{Magnetar Engine}
\label{sec:magnetar}

Here we explore the possibility that the long duration light curve is powered by the spin-down energy of a millisecond magnetar \citep{kasen10}. This model has been successful at explaining the diverse light curves of Type~I SLSNe (see \citealt{nicholl17} for a detailed description of the model and choice of priors).  We note that this model cannot account for the double-peaked structure of the light curve, but modest modulations have been previously observed in SLSNe \citep{leloudas12,nicholl15,smith16,vreeswijk17,blanchard18}.

The most likely magnetar models are shown in the top panel of Figure~\ref{fig:magnetar}, and the model parameters are listed in Table~\ref{tab:magnetar}, where we use the observed width of the emission lines as a prior for the velocity of the ejecta. The model requires an initial rapid spin period of \mbox{$P\approx 1.5$ ms}, a magnetic field of \mbox{$B_{\perp}\approx 3\times 10^{14}$ G}, and a kinetic energy of $E_K\approx 8\times 10^{51}$ erg. These values are within the typical range inferred for SLSNe: \mbox{$P\approx 1-5$ ms}, \mbox{$B_{\perp}\approx (0.1 - 3)\times 10^{14}$G}, and  \mbox{$E_K \approx (1 - 30)\times 10^{51}$ erg s$^{-1}$} \citep{nicholl17}. However, this model requires a large ejecta mass of \mbox{$\approx 90$ M$_\odot$}, much higher than the typical values of \mbox{$1-20$ M$_\odot$} for SLSNe \citep{nicholl17}. We note that the magnetar model also nicely accounts for the late-time flattening of the light curves, but as mentioned above, does not strictly account for the double-peaked light curve structure.

A key challenge to the magnetar model, however, is that the inferred large ejecta mass falls in the regime for PISNe (i.e., helium core mass of \mbox{$60-137$ M$_\odot$}; \citealt{heger02}), which is inconsistent with the required neutron star remnant.  About $1\%$ of the posterior probability on $M_{\rm ej}$ resides at masses below the PISN regime (i.e., in the PPISN regime) but in this case too the results are inconsistent since the star is expected to shed about half of its mass in the pulsational phase prior to explosion \citep{woosley17}.

\begin{deluxetable*}{ccccc}
	\tablecaption{Best-fit Parameters for the Magnetar and Fallback Central Engine Models. \label{tab:magnetar}}
	\tablehead{\colhead{Parameter} & \colhead{Prior}  & \colhead{Magnetar}  & \colhead{Fallback}  & \colhead{Units}}
	\startdata
	$M_{\text{ej}}                $ & $[1, 260]$         & $ 91^{+24}_{-17}       $ & $ 93^{+33}_{-24}        $ & M$_\odot$              \\
	$\log(L_1)                    $ & $[51, 57]$         &   \nodata                & $ 55.2^{+0.2}_{-0.3}  $ & erg s$^{-1}$           \\
	$\log{(t_{\text{on}})}        $ & $[-4, 2]$          &   \nodata                & $ -2.1 \pm 1.3          $ & days                   \\
	$M_{\text{NS}}                $ & $[1.4, 2.2]$       & $ 1.9^{+0.2}_{-0.3}    $ &   \nodata                 & M$_\odot$              \\
	$P_{\text{spin}}              $ & $[0.7, 20]$        & $ 1.5 \pm 0.4          $ &   \nodata                 & ms                     \\
	$B_{\perp}                    $ & $[0.01, 10]$       & $ 3.2 ^{+0.9}_{-0.7}   $ &   \nodata                 & $10^{14}$ G            \\
	$V_{\text{ej}}                $ & $3400 \pm 900$     & $ 3699^{+492}_{-476}   $ & $ 3783^{+797}_{-683}    $ & km s$^{-1}$            \\
	$E_{k}^{\dagger}              $ &                    & $ 7.5^{+3.5}_{-2.4}    $ & $ 8.1^{+6.2}_{-3.8}    $ & $10^{51}$ erg s$^{-1}$ \\
	$t_{\text{exp}}               $ & $[0, 500]$         & $ 56.9^{+8.6}_{-10.0}  $ & $ 66.1^{+11.8}_{-12.1}    $ & days                   \\
	$T_{\text{min}}               $ & $[1000, 15000]$    & $ 7648^{+395}_{-270}   $ & $ 7821^{+486}_{-373}    $ & K                      \\
	$\log{(n_{H,\text{host}})}    $ & $[16, 23]$         & $ 18.6 \pm 1.8         $ & $ 18.5 \pm 1.7   $ & cm$^{-2}$              \\
	$A_{V, \text{host}}^{\dagger} $ &                    & $ < 0.1              $ & $ < 0.08          $ & mag                    \\
	$\kappa                       $ & $[0.1, 0.2]$       & $ 0.16 \pm 0.03        $ & $  0.14 \pm 0.04               $ &                        \\
	$\log{(\kappa_{\gamma})}      $ & $[-4, 4]$          & $ 1.5 \pm 1.7          $ &   $  -2.3 \pm 0.15               $             & cm$^2$g$^{-1}$         \\
	$\log{\sigma}                 $ & $[-4, 2]$          & $-0.75 \pm 0.04        $ & $-0.75 \pm 0.03         $ &                        \\
	\enddata
	\tablecomments{Best model parameters, prior ranges, and 1$\sigma$ error bars for the realizations shown in Figure~\ref{fig:magnetar}. See Table~\ref{tab:parameters} for parameter definitions.}
	\tablenotetext{\dagger}{These parameters were not fit for, but were calculated using all the posterior distribution samples of the fitted parameters.}
\end{deluxetable*}

\subsection{Fallback Accretion}
\label{sec:fallback}

As an alternative central engine, we model the light curve with a fallback accretion model, in which some of the explosion debris remains bound and accretes onto the compact remnant \citep{dexter13}. The {\tt MOSFiT} implementation of this model is described in \cite{moriya18}. Our model is parametrized in a way that takes the engine power input $L_1$ (The luminosity at 1 second, assuming a power law extrapolation), and calculates a constant input luminosity \mbox{$L_{\rm on} = L_1 t_{\rm on}^{-5/3}$} up to time $t_{\rm on}$, after which the input luminosity declines at a rate \mbox{$L\propto t^{-5/3}$}. The best fit model for SN\,2016iet has \mbox{$L_1 = 10^{55.2}$ erg s$^{-1}$} and \mbox{$t_{\rm on}\approx 690$ s}, which corresponds to an initial input luminosity of \mbox{$L_{\rm on}\approx 3\times10^{50}$ erg s$^{-1}$} during the first 690~s.  We can estimate the total fallback mass from the accretion rate as:
\begin{equation}
M_{\rm fb} = \frac{L_1}{c^2\epsilon} \left[ \int_{0}^{t_{\rm on}} t_{\rm on}^{-5/3}dt + \int_{t_{\rm on}}^{t_f} t^{-5/3}dt \right].
\end{equation}
Assuming an efficiency of $\epsilon = 0.1$, and integrating for \mbox{$t_f = 760$ days}, this yields $M_{\rm fb}\approx 1.7$ M$_\odot$. 

The most likely fallback accretion models are shown in the bottom panel of Figure~\ref{fig:magnetar}, and the model parameters are listed in Table~\ref{tab:magnetar}. Overall, since the energy input rate is similar to that of a magnetar engine (i.e., $t^{-5/3}$ vs.~$t^{-2}$), the resulting light curves and model parameters are nearly identical for the two models.  In particular, we find \mbox{$M_{\rm ej}\approx 93$ M$_\odot$}. This model can also account for the late-time flattening of the light curves, but not for the double-peaked structure.

This model therefore suffers from the same problems as the magnetar central engine model. Namely, the inferred progenitor mass lies in the PISN regime, expected to leave behind no remnant. Only about $6\%$ of the posterior probability on $M_{\rm ej}$ is in the PPISN regime, which still leads to an inconsistency, as discussed in the previous section.

\begin{figure}
	\begin{center}
		\includegraphics[width=0.96\columnwidth]{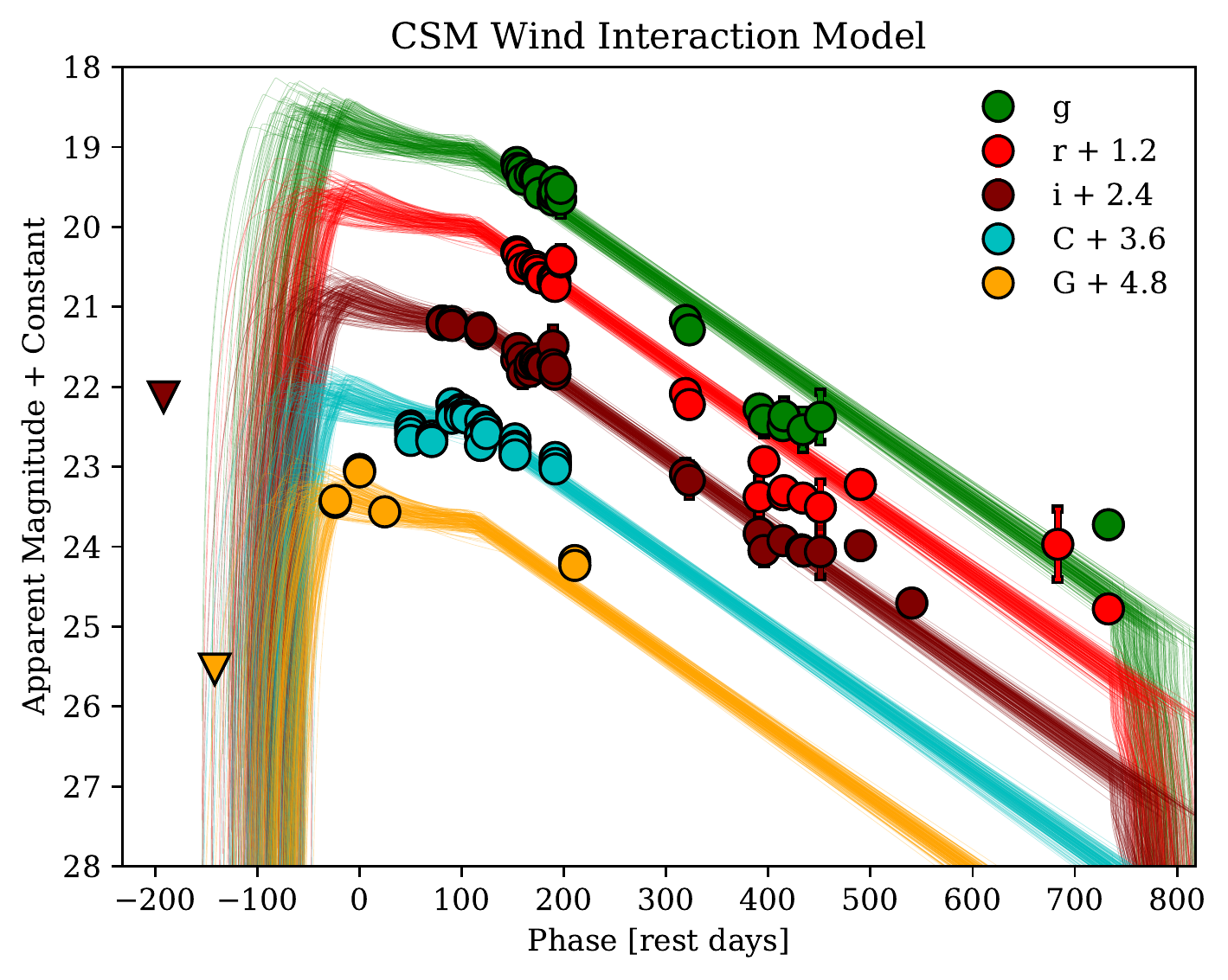}
		\includegraphics[width=0.96\columnwidth]{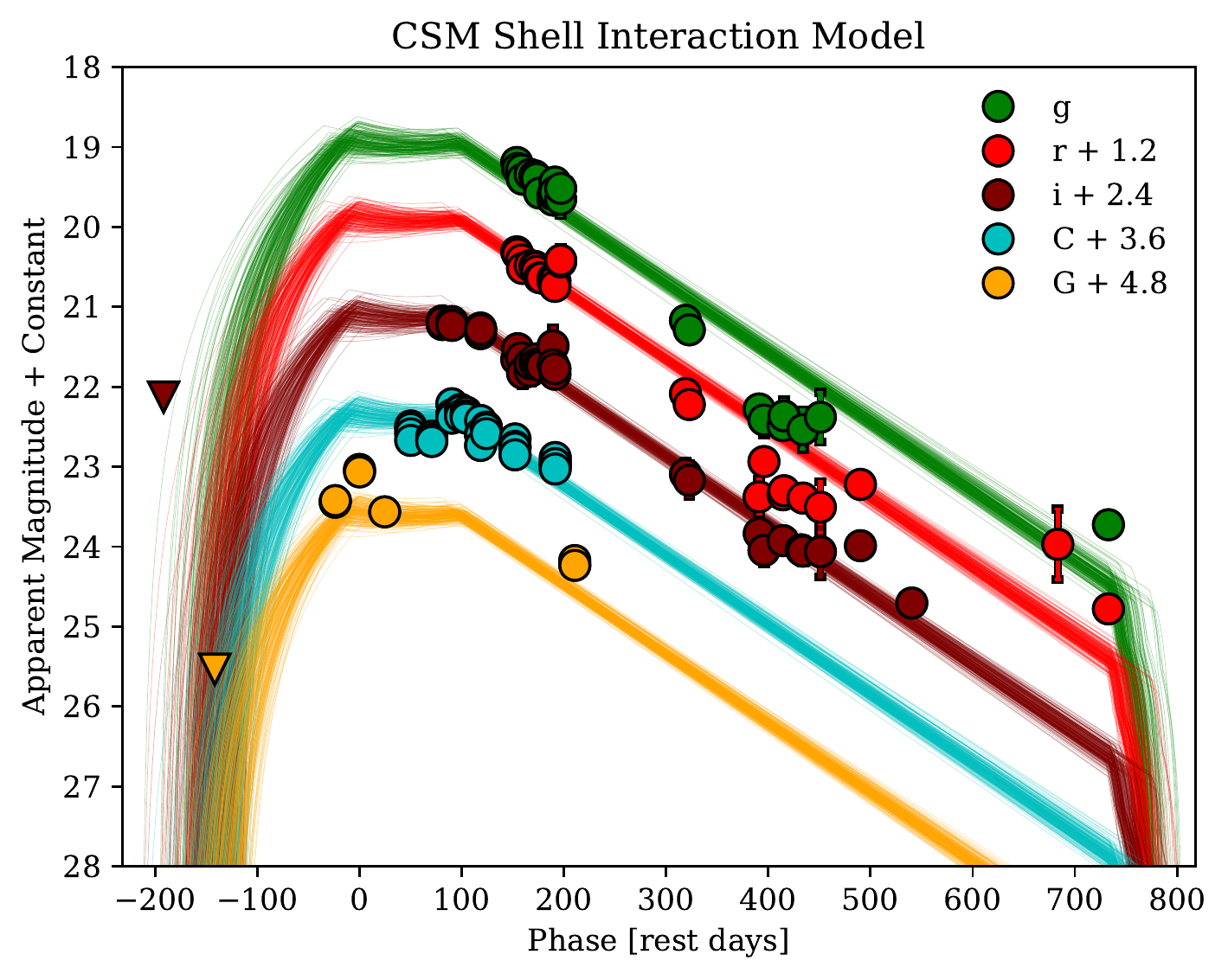}
		\includegraphics[width=0.96\columnwidth]{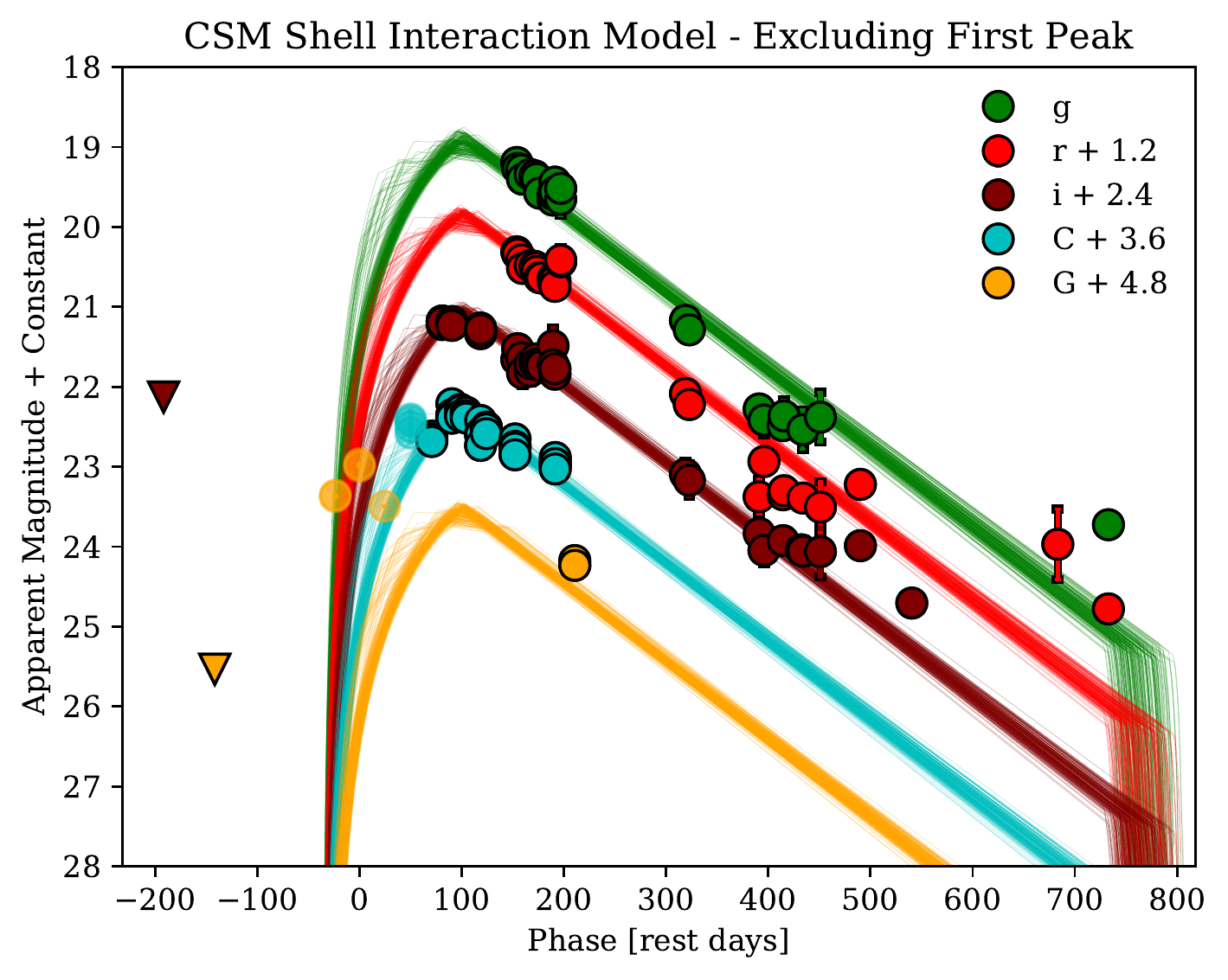}
		\caption{Light curve of SN\,2016iet with CSM interaction models for a wind density profile ($\rho \propto r^{-2}$; {\it Top}), a constant density shell profile ({\it Middle}); and a shell profile fit to only the second peak and beyond ({\it Bottom}). Each line is a sample realization of the most likely models generated from {\tt MOSFiT}. The resulting parameters are provided in Table~\ref{tab:csm}, and the models are discussed in \S\ref{sec:csm} and \S\ref{sec:shock}. \label{fig:csm}}
	\end{center}
\end{figure}

\subsection{CSM Interaction} 
\label{sec:csm}

We next fit the light curve with the CSM interaction model described in \citet{chatzopoulos13}, where the radiation is produced by the interaction of the SN ejecta with a CSM that has either a wind ($\rho\propto r^{-2}$) or a shell (constant density) profile. In this model, the double-peaked structure of the light curve could in principle be explained as being due to the forward and reverse shocks. We use the inferred radius of the photosphere (calculated in \S\ref{sec:bolometric}) to constrain the inner radius of the CSM, $R_0$, such that the radius of the photosphere is located inside the CSM at all times (i.e., $R_0<R_{\rm ph}$). For this reason, we set the largest value of $R_0$ to 54 AU, the smallest robustly measured photospheric radius at a phase of 350 days.

The results of the wind and shell interaction models are shown in the top and middle panels of Figure~\ref{fig:csm}, and the full set of parameters is provided in Table~\ref{tab:csm}. In both models we find a large CSM mass of \mbox{$M_{\rm CSM}\approx 37$ M$_\odot$}.  The ejecta masses are similarly large, \mbox{$M_{\rm ej}\approx 85$} and \mbox{$\approx 45$ M$_\odot$} for the wind and shell models, respectively.  The ejecta velocities of \mbox{$v_{\rm ej}\approx 4100$} and \mbox{$\approx 2150$ km s$^{-1}$} in the wind and shell models are comparable to the observed line widths, although the value for the wind model is a better match (this is not influenced by the choice of prior). The inferred inner radius of the CSM is influenced by our choice of prior, with values of \mbox{$R_0\approx 43$} and \mbox{$\approx 26$ AU}, respectively.

The CSM models are able to roughly capture the double-peaked structure of the light curve as the forward and reverse shocks. The models do not explain the late-time flattening, although this could be interpreted as the signature of an increase in the density of the CSM at larger radii relative to the simple density profiles assumed here. Finally, we note that the CSM interaction model is expected to produce a blue continuum, which we indeed observe in the spectra (Figure~\ref{fig:spectra}).  Thus, the CSM models provide an adequate explanation for the light curves and spectra, and require a progenitor mass shortly before explosion (i.e., a CO core mass of $M_{\rm ej}+M_{\rm CSM}$) of $\approx 85-120$ M$_\odot$, squarely in the PISN regime.

\begin{deluxetable*}{cccccc}
	\tablecaption{Best-fit Parameters for the CSM Interaction Models. \label{tab:csm}}
	\tablehead{\colhead{Parameter} & \colhead{Prior}  & \colhead{Wind CSM}  & \colhead{Shell CSM} & \colhead{Shell CSM Excluding First Peak} & \colhead{Units}}
	\startdata
	$M_{\text{ej}}                $ & $[1, 260]$         & $ 84.6^{+18.3}_{-15.3}  $ & $ 45.4^{+15.9}_{-7.3}    $ & $ 18.9^{+10.0}_{-8.7}    $ & M$_\odot$              \\
	$M_{\text{CSM}}               $ & $[1, 260]$         & $ 36.3^{+8.3}_{-7.5}    $ & $ 37.9^{+12.6}_{-6.5}    $ & $ 33.0^{+6.4}_{-7.5}    $ & M$_\odot$              \\
	$V_{\text{ej}}                $ & $[1000, 10000]$    & $ 4100^{+505}_{-412}   $ & $ 2147^{+342}_{-198}    $ & $ 2599^{+539}_{-308}    $ & km s$^{-1}$            \\
	$E_{k}^{\dagger}              $ &                    & $ 8.7^{+1.3}_{-1.5}     $ & $ 1.2 ^{+0.4}_{-0.1}         $ & $ 0.7 ^{+0.5}_{-0.1}          $ & $10^{51}$ erg s$^{-1}$ \\
	$R_0                          $ & $[1, 54.2]^a$      & $ 42.8^{+8.2}_{-11.1}          $ & $ 26.3^{+19.8}_{-18.8}   $ & $ 44.4 \pm 7.0          $ & AU                     \\
	$\log(\rho)                   $ & $[-14, -9]$        & $ -11.1 \pm 0.2         $ & $ -12.0 \pm 0.2         $ & $ -11.8 \pm 0.2         $ & g cm$^{-3}$            \\
	$t_{\text{exp}}               $ & $[0, 500]^a$       & $ 60.0^{+15.8}_{-26.5}   $ & $ 128 ^{+15}_{-19}          $ & $ 107.1^{+4.4}_{-4.7}    $ & days                   \\
	$T_{\text{min}}               $ & $[1000, 15000]$    & $ 7753^{+416}_{-334}   $ & $ 7889^{+973}_{-407}    $ & $ 7652^{+455}_{-306}    $ & K                      \\
	$\log{(n_{H,\text{host}})}    $ & $[16, 23]$         & $ 18.2^{+1.9}_{-1.6}     $ & $ 18.7^{+2.1}_{-1.9}     $ & $ 18.6 \pm 1.8     $ & cm$^{-2}$              \\
	$A_{V, \text{host}}^{\dagger} $ &                    & $ < 0.06                $ & $ < 0.30                $ & $ < 0.13                $ & mag                    \\
	$\kappa                       $ & $[0.1, 0.2]$       & $ 0.14^{+0.04}_{-0.03}  $ & $ 0.14^{+0.04}_{-0.03}  $ & $ 0.13^{+0.04}_{-0.02}  $ &                        \\
	$\log{\sigma}                 $ & $[-4, 2]$          & $-0.63 \pm 0.04         $ & $-0.61 \pm 0.04         $ & $ -0.75 \pm 0.04        $ &                        \\
	\enddata
	\tablecomments{Best model parameters, prior ranges, and 1$\sigma$ error bars for the realizations shown in Figure~\ref{fig:csm}. See Table~\ref{tab:parameters} for parameter definitions. The maximum on the prior of $R_0$ is motivated by the fact that the photosphere radius should be greater than the inner radius of the CSM.}
	\tablenotetext{\dagger}{These parameters were not fit for, but were calculated using all the posterior distribution samples of the fitted parameters.}
	\tablenotetext{a}{Priors only used for Wind and Shell CSM models. The priors for the ``Shell CSM Exclusing First Peak" model were set to $[34.65,54.2]$ AU and $[113.7, 100.9]$ days, respectively.}
\end{deluxetable*}

\subsection{Shock Breakout Plus CSM Interaction}
\label{sec:shock}

Finally, we explore a variant of the CSM interaction model in which the first peak is due to shock breakout from an inner dense CSM and the second peak and subsequent light curve are powered by CSM interaction as in the previous section. 

\begin{figure}
	\begin{center}
		\includegraphics[width=\columnwidth]{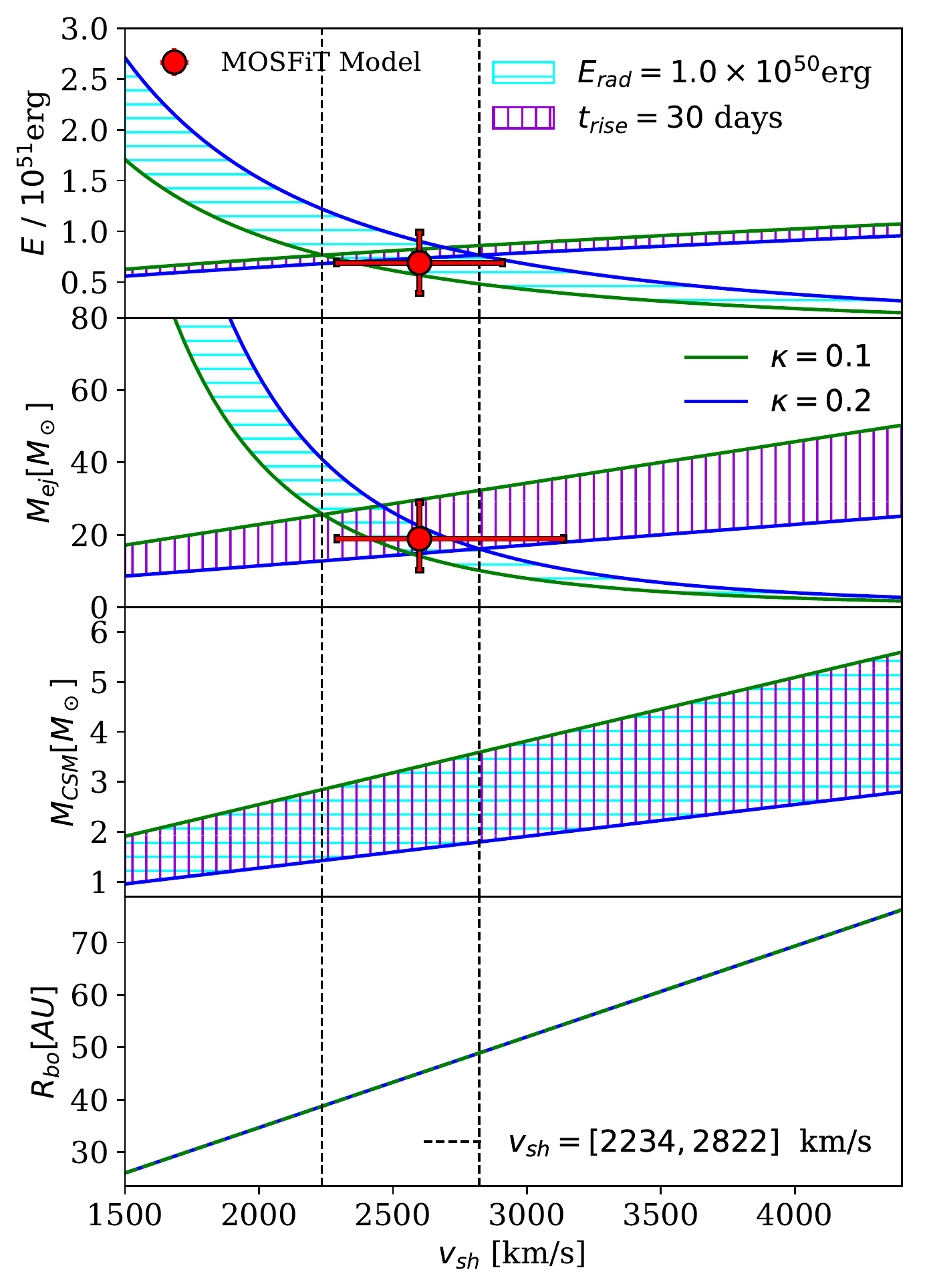}
		\caption{Solutions to the shock breakout equations of \cite{chevalier11} used to calculate the kinetic energy $E_{K,51}$, ejecta mass $M_{\rm ej}$, CSM mass $M_{\rm CSM}$, and breakout radius $R_{\rm bo}$ from the inferred radiated energy $E_{\rm rad}$, rise time $t_{\rm rise}$, opacity $\kappa$, and shock velocity $v_{\rm sh}$. The cyan band represents the solution with a fixed radiated energy of $10^{50}$ erg, and the purple band represents the solution with a fixed rise time of 30 days. The regions where both bands overlap are the allowed solutions. The green and blue boundaries are determined by the limits in the allowed opacity $\kappa = [0.1, 0.2]$. The red data points are the result from the {\tt MOSFiT} CSM interaction models excluding the first peak shown in the bottom panel of Figure~\ref{fig:csm}. We find that the two fits are fully self consistent. \label{fig:shock}}
	\end{center}
\end{figure}

We model the first peak with the shock breakout model of \citet{chevalier11} (Equations 3 and 5 in that paper) from which we can infer the breakout radius $R_{\rm bo}$, as well as $M_{\rm ej}$, $M_{\rm CSM}$, and $E_K$.  The input parameters are the light curve rise time $t_{\rm rise}$, the radiated energy $E_{\rm rad}$, the shock velocity $v_{\rm sh}$, and an optical opacity $\kappa$. We estimate $t_{\rm rise}\sim 30$ days from a second-order polynomial fit to the first peak, which is similar to the 28 days separation between the first and brightest \textit{Gaia} data points.  We calculate \mbox{$E_{\rm rad}\sim L_{\rm p}\times t_{\rm rise}\sim 10^{50}$ erg}, where \mbox{$L_{\rm p}\sim 10^{43}-10^{44}$ erg s$^{-1}$} (calculated in \S\ref{sec:bolometric}). We set the value of $\kappa$ to $0.1-0.2$, typical values assumed for hydrogen poor material. The resulting ejecta and CSM parameters as a function of $v_{\rm sh}$ are shown in Figure~\ref{fig:shock}.  We find that self-consistent solutions require $v_{\rm sh}\approx 2200-2800$ km s$^{-1}$, comparable to the observed emission line widths in the spectra.  The inferred breakout radius is \mbox{$R_{\rm bo}\approx 40-50$ AU}, the ejecta mass is \mbox{$M_{\rm ej}\approx 15-28$ M$_\odot$}, and the CSM mass of the breakout region is $\approx 1-4$ M$_\odot$. We find a parameter $D_* = 8-16$ \citep{chevalier11}, depending on the assumed opacity, corresponding to a mean density in the breakout region of \mbox{$\rho\sim 2\times 10^{-11}$ g cm$^{-3}$}.

\begin{figure}[t]
	\begin{center}
		\includegraphics[width=\columnwidth]{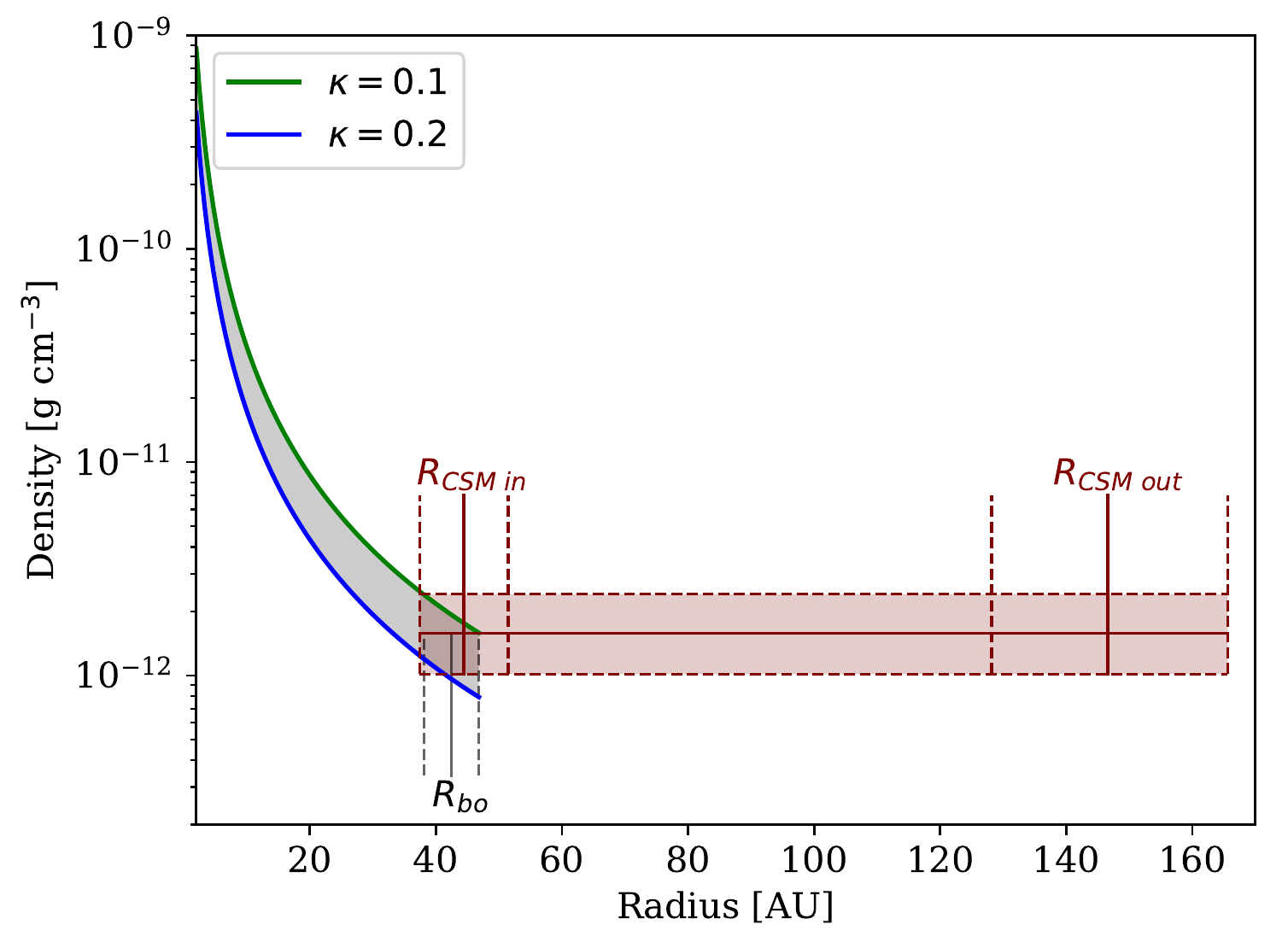}
		\caption{Inferred density profile of the CSM from the shock breakout model of \citet{chevalier11} for the first peak of the light curve (grey band), and a shell density profile CSM model for the second part of the light curve (red band; bottom panel of Figure~\ref{fig:csm}).  The two density profiles are fully consistent. \label{fig:density}}
	\end{center}
\end{figure}

We model the rest of the light curve with interaction with a shell-like CSM as in \S\ref{sec:csm}, but with an inner radius \mbox{$R_0\gtrsim 35$ AU}, the smallest inferred value of $R_{\rm bo}$, shown in Figure~\ref{fig:density}. We find \mbox{$M_{\rm CSM}\approx 33$ M$_\odot$}, \mbox{$M_{\rm ej}\approx 19$ M$_\odot$}, and \mbox{$v_{\rm ej}\approx 2600$ km s$^{-1}$}; see the bottom panel of Figure~\ref{fig:csm} and Table~\ref{tab:csm} for the model light curve and parameter values. Comparing these results to the breakout analysis of the first peak we find consistent values of $v_{\rm ej}$, $M_{\rm ej}$ and $E_K$.  Considering the overall simplified breakout model and the rough estimate of the rise time and radiated energy of the first peak we consider the overall agreement to indicate that this is a plausible interpretation of the double-peaked structure.

This result is also supported by the density inferred from the two components.  In Figure~\ref{fig:density} we plot the density profile from the breakout model, which following \citet{chevalier11} we assume to be a wind profile, and the constant density inferred from the CSM interaction model of the second peak. We find that the two join smoothly at \mbox{$R_{\rm bo}\approx R_{\rm CSM,in}\approx 40$ AU} with a density of $\sim 10^{-12}$ g cm$^{-3}$.

\begin{figure}
	\begin{center}
		\includegraphics[width=\columnwidth]{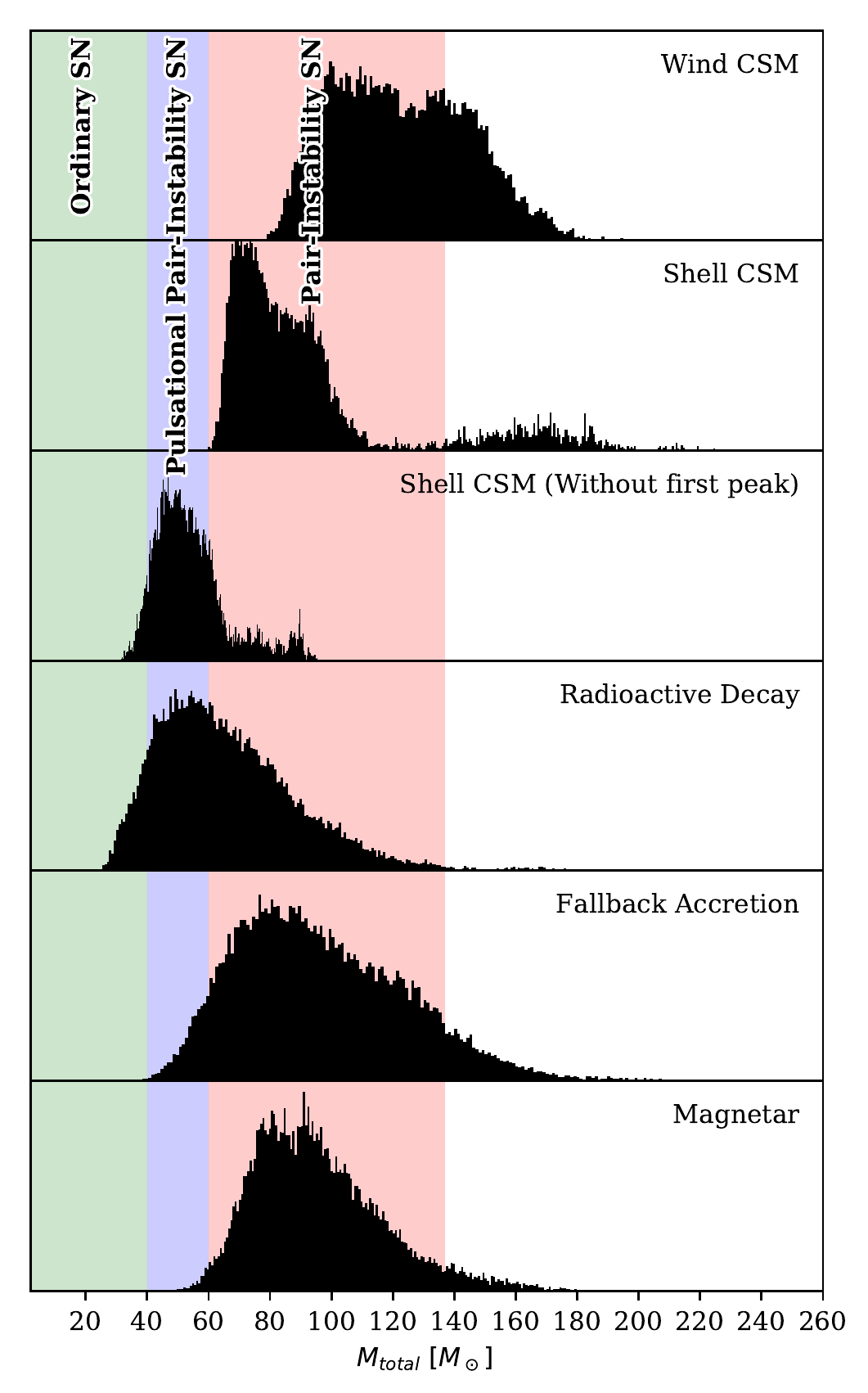}
		\caption{Inferred ejecta or ejecta+CSM masses from the different model fits to the light curve. The shaded regions mark the theoretical ranges for core-collapse SNe (green), PPISNe (blue), and PISNe (pink).  The posterior mass distributions for the models of SN\,2016iet all overlap the PPISN or PISN ranges. \label{fig:masses}}
	\end{center}
\end{figure}

\subsection{Summary of the Light Curve Models}

We modeled the light curves of SN\,2016iet with a broad range of energy sources.  We find that a radioactive decay model fit to the entire light curve requires a large ejecta mass of \mbox{$\approx 64$ M$_\odot$} and a nickel mass of \mbox{$\approx 3.7$ M$_\odot$}.  This model has several disadvantages; it cannot account for the double-peaked structure or the late-time flattening of the light curves, it severely underestimates the latest measurements, and it requires a nickel mass fraction that is not supported by the spectroscopic data.  We therefore consider this model unlikely.  A radioactive model fit for just the first peak requires an unusually large nickel fraction ($\sim 0.24$), making it an unlikely explanation. While the latter possibility cannot be ruled out, it still requires an explanation for the rest of the light curve.

The magnetar and fallback accretion central engine models both require a large ejecta mass of \mbox{$\approx 90$ M$_\odot$}. These models do not explain the double-peaked light curve structure, but they nicely capture the late-time flattening.  The inferred engine parameters for the magnetar central engine are similar to those of SLSNe, but the ejecta mass is much larger. A key theoretical problem for the engine models, however, is that progenitors with a helium core mass this high are expected to produce PISNe, which do not leave behind a neutron star or black hole remnant.

In the ejecta-CSM interaction models, the double-peaked structure is  accommodated by emission from the forward and reverse shocks.  This model can also account for the blue continuum seen in the spectra in the first year. The resulting CSM mass is \mbox{$\approx 37$ M$_\odot$} and the ejecta mass is \mbox{$M_{\rm ej}\approx 45-85$ M$_\odot$}, depending on the density profile of the CSM.  Therefore, the progenitor had an inferred mass of \mbox{$\approx 85-120$ M$_\odot$} shortly before explosion, after losing its hydrogen and helium layers. This corresponds to a progenitor ZAMS mass of \mbox{$\approx 150 - 260$ M$_\odot$}, the regime expected to produce PISNe. The late-time flattening is not captured in these models, but it can in principle be attributed to a change in the density profile at larger radii. 

In the breakout+interaction model, we find an ejecta mass of \mbox{$\approx 19$ M$_\odot$}, an inner breakout region with a mass of \mbox{$\approx 3$ M$_\odot$} extending to about 40 AU, and an outer CSM with a mass of \mbox{$\approx 33$ M$_\odot$}.  This model therefore requires the lowest progenitor mass, with a CO core of \mbox{$\approx 55$ M$_\odot$}. This variation of the CSM model provides a consistent result in terms of the ejecta and shock velocity, ejecta mass, and kinetic energy from the breakout-powered first peak and the interaction-powered second peak (Figure~\ref{fig:shock}). 

The implication of the CSM models (regardless of variant) is a mass of \mbox{$M_{\rm CSM}\approx 35$ M$_\odot$} located between inner and outer radii of \mbox{$\approx 40-150$ AU} (e.g., Figure~\ref{fig:density}).  For an ejection velocity of \mbox{$\approx 100$ km s$^{-1}$}, this requires a mean mass loss rate of \mbox{$\dot{M}\approx 7$ M$_\odot$ yr$^{-1}$} spanning the final \mbox{$\approx 2-7$ years} before explosion.  If the ejection velocity was instead \mbox{$1000$ km s$^{-1}$} the corresponding timescale over which the CSM was ejected would be only $70-260$ days before explosion, with a mean rate of \mbox{$\approx 0.2$ M$_\odot$ day$^{-1}$}.

The key result from our modeling is that regardless of the model, the inferred total mass (ejecta or ejecta+CSM) is well outside the range inferred for normal Type~I SNe, including the most energetic Type~Ic events and SLSNe, which have $M_{\rm ej}\lesssim 20$ M$_\odot$ \citep{modjaz16,nicholl17}.  We show the model posterior distributions of the total mass in Figure~\ref{fig:masses} in comparison to theoretical predictions for various explosion mechanisms. Stars with a helium core mass of $2-40$ M$_\odot$ are expected to produce an ordinary core-collapse SN, or in some cases a failed SN. Stars with a helium core mass of $40-60$ M$_\odot$ are expected to produce a PPISNe, while those with a helium core mass of $60-137$ M$_\odot$ are expected to produce a PISNe \citep{woosley07}. All of the models for SN\,2016iet place the progenitor in the PPISN or PISN regimes, especially the CSM models, which provide the most consistent explanations for the light curves and spectra.   

\section{Host Galaxy} 
\label{sec:host} 

\begin{figure}[t]
	\begin{center}
		\includegraphics[width=\columnwidth]{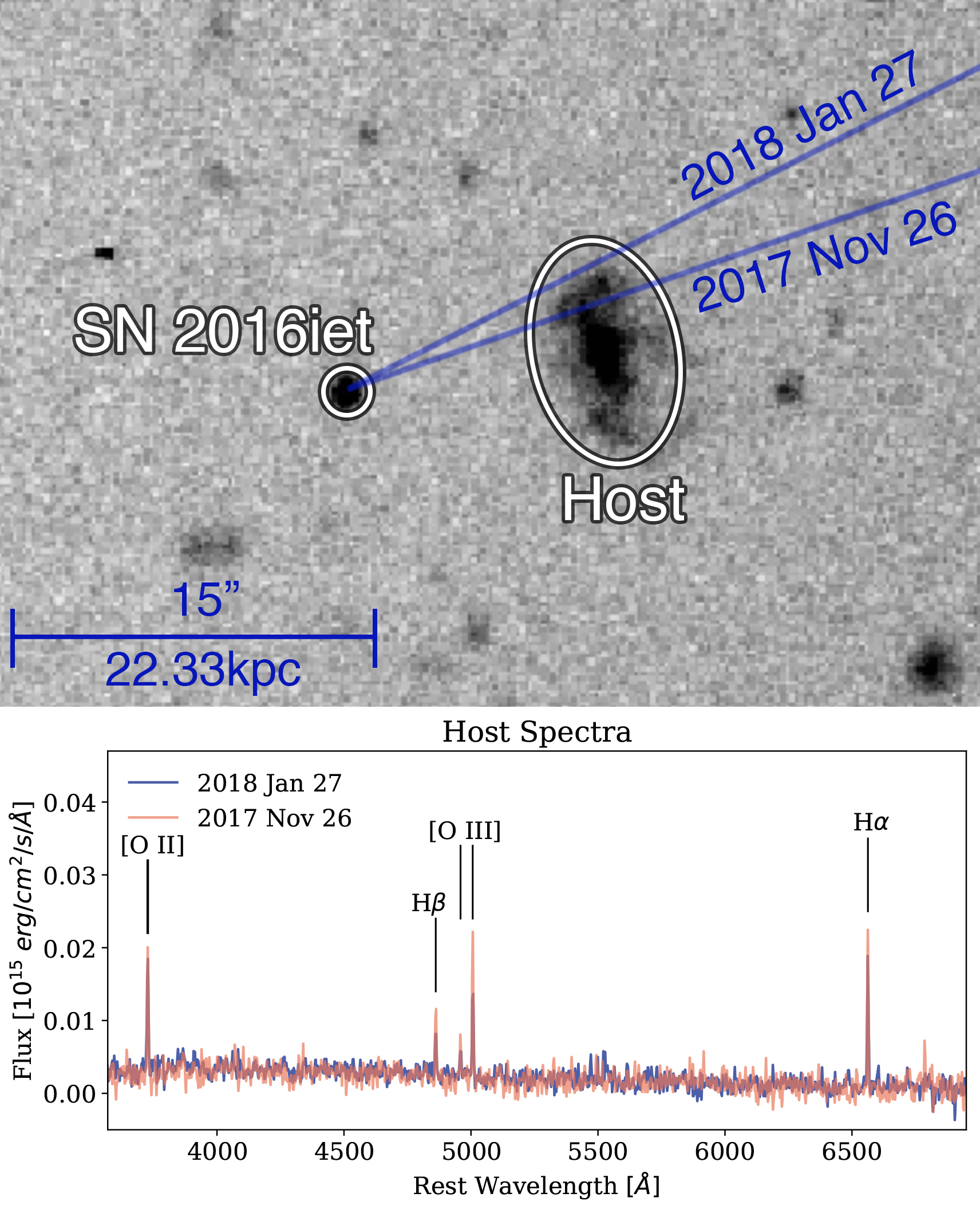}
		\caption{Image of SN\,2016iet and its most likely host galaxy taken with the LDSS3c instrument in $i$-band on 2018 July 9. The blue lines mark the position of the slit during two spectral observations that included the host. The resulting host spectra are shown in the lower panel. \label{fig:host_image}}
	\end{center}
\end{figure}

No galaxy is detected at the location of SN\,2016iet in archival PS1/$3\pi$ images to $3\sigma$ limits of $g\gtrsim 23.1$ mag, $r\gtrsim 23.0$ mag, $i\gtrsim 22.7$ mag, $z\gtrsim 22.0$ mag, and $y\gtrsim 21.0$ mag. The nearest galaxy with a low probability of chance coincidence of $P_{\rm cc}\approx 0.1$, is located \mbox{$11.1\pm 0.3''$} from the SN position (Figure~\ref{fig:host_image}).  We measure a redshift for this galaxy of \mbox{$z = 0.0671$ $\pm 0.0001$} from several prominent narrow emission lines (H$\alpha$, H$\beta$, \mbox{[\ion{O}{2}] $\lambda\lambda$3726,3729}, and \mbox{[\ion{O}{3}] $\lambda\lambda$4959,5007}; see Figure~\ref{fig:host_image}). This represents a velocity difference of \mbox{$140\pm 85$ km s$^{-1}$} relative to the redshift inferred from the emission lines in SN\,2016iet. We therefore conclude that SN\,2016iet is associated with this galaxy.

We calculate a projected physical separation of SN\,2016iet to this host galaxy of \mbox{$16.5\pm 0.6$ kpc}. Relative to the half-light radius of the host galaxy, \mbox{$R_e= 3.8\pm 0.4$ kpc}, this corresponds to a normalized offset of \mbox{$D/R_e=4.3\pm 0.4$}. \citet{kelly12} found that Type~Ic SNe tend to occur close to their host galaxies, with $D/R_e\lesssim 2.4$. Similarly, SLSNe exhibit offsets of $D/R_e\lesssim 1.6$ \citep{lunnan15}, while long GRBs have normalized offsets of $D/R_e\lesssim 1.8$, where only one out of a sample of 105 has a separation $D/R_e\approx 4$ \citep{blanchard16}. Calcium rich gap transients show comparable host offsets, with a median of $D/R_e\approx 10$ \citep{lunnan17}, but these transients tend to occur in massive galaxies \citep{lunnan17,de2018}, unlike the host of SN\,2016iet. In general, massive progenitors tend to occur at small offsets from their hosts, whereas SN\,2016iet clearly violates this trend. 

We extract the host's magnitudes by performing photometry on stacked PS1/$3\pi$ $grizy$ template images using a constant circular aperture, to ensure uniform sampling in all filters. We include additional $ugr$ measurements from our IMACS and LDSS3c images. The resulting magnitudes are listed in Table~\ref{tab:host}, along with the corresponding absolute magnitudes.  We find $M_r\approx -17.5$ mag, or $L_r\approx 0.02\, L_*$ \citep{montero09}, identical to that of the SMC. If SN\,2016iet instead originated in an underlying satellite galaxy, then the brightness of this galaxy would be $M_r\gtrsim -13.9$ mag, or $L_r\lesssim 0.003\, L_*$.

\begin{figure}
	\begin{center}
		\includegraphics[width=\columnwidth]{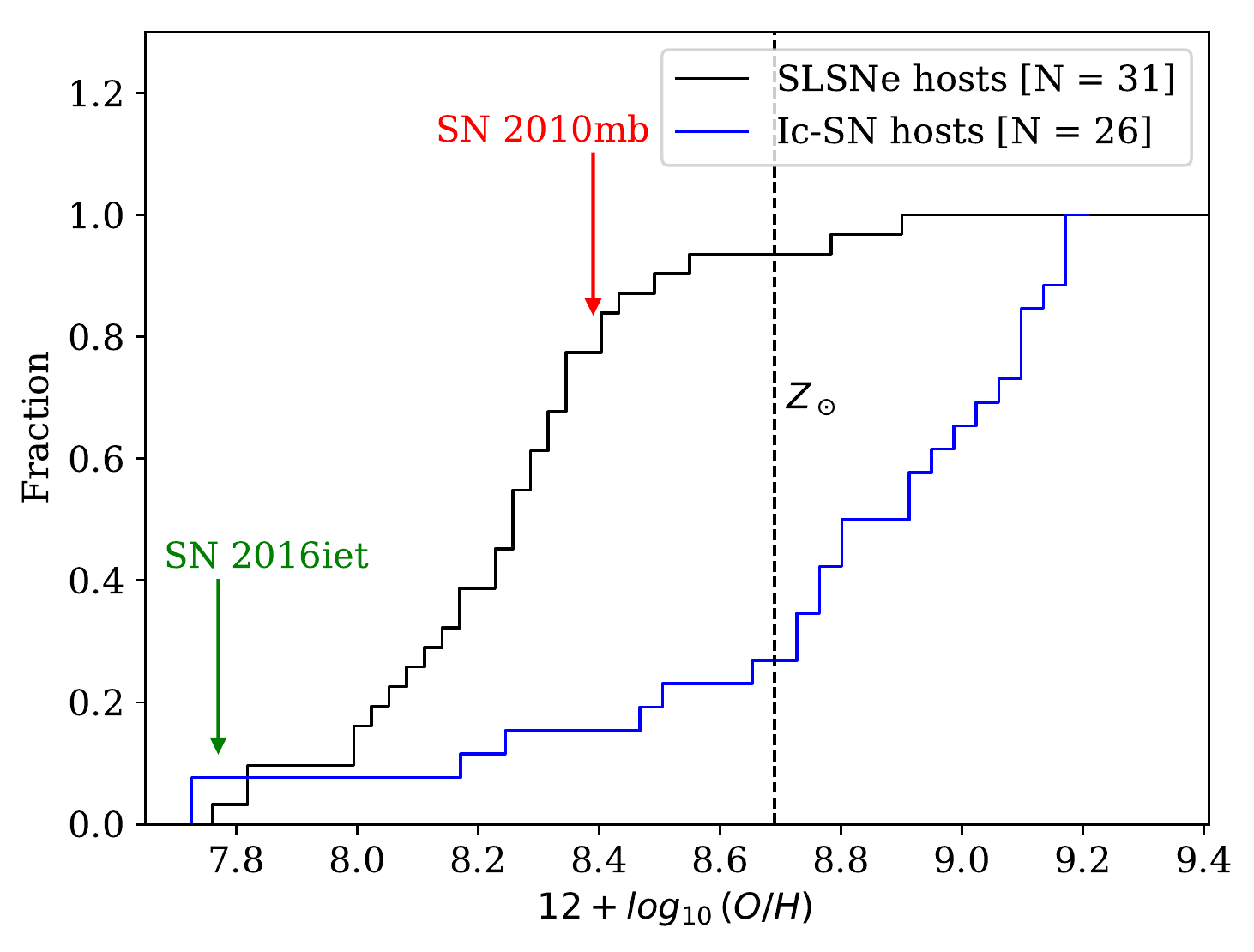}
		\caption{The inferred metallicity of the host galaxy of SN\,2016iet (green) relative to the cumulative distribution of the host galaxy metallicities for
			Type~Ic SNe \citep{modjaz19} and SLSNe \citep{lunnan14}.  Also shown in red is the host metallicity for SN\,2010mb \citep{benami14}. \label{fig:metal}}
	\end{center}
\end{figure}

The flux ratio of \mbox{$L_{\rm H\alpha}/L_{\rm H\beta}=2.3\pm 0.6$} is consistent with the expectation for case B recombination ($L_{\rm H\alpha}/L_{\rm H\beta} = 2.86$), indicating no evidence for significant host galaxy extinction. We infer the metallicity using the $R_{23}$ method \citep{kobulnicky99}, which is double-valued, but the lack of detectable [\ion{N}{2}] emission and the low luminosity of the host galaxy point towards the lower branch solution. This gives a value of \mbox{$12 + \log($O/H$) = 7.80^{+0.15}_{-0.12}$}, or \mbox{$Z = 0.14^{+0.05}_{-0.03}$ Z$_\odot$} \citep{asplund09}. Given the large offset, the metallicity at the location of SN\,2016iet is likely even lower. We stress that this metallicity is low even in comparison to the host galaxies of SLSNe, which occur preferentially in low metallicity galaxies, but mostly span $12+\log($O/H$)\approx 7.8-8.6$ \citep{lunnan14}; see Figure~\ref{fig:metal}. Similarly, only 5\% of the Type~Ic SN sample from \citet{modjaz19} have metallicities lower than the value inferred for SN\,2016iet, namely SNe 2010hie and 2011rka.

\begin{deluxetable*}{ccc|ccc}
	\tablecaption{Host Galaxy Properties \label{tab:host}}
	\tablehead{\colhead{}           & \colhead{Value}           & \colhead{Units} & \colhead{}           & \colhead{Value}           & \colhead{Units}}
	\startdata
    $P_{\rm cc}$   & 0.12                 &              &  $M_z$                       & $-17.50 \pm 0.12$        & mag                     \\
	$z$            & $0.0671 \pm 0.0001$  &              &  $M_y$                       & $-17.35 \pm 0.34$        & mag                     \\
	$D$            & $11.1\pm 0.3$        & arcsec       &  $L_u$                       & $0.28 \pm 0.04$          & $L_*$                   \\
	$D$            & $16.5\pm 0.6$        & kpc          &  $L_g$                       & $0.076 \pm 0.004$        & $L_*$                   \\
	$r_{50}$       & $2.57 \pm 0.26$      & arcsec       &  $L_r$                       & $0.0234 \pm 0.0004$      & $L_*$                   \\
	$v_r$          & $140\pm 85$          & km s$^{-1}$  &  $L_i$                       & $0.020 \pm 0.002$        & $L_*$                   \\
	$u$            & $20.27 \pm 0.15$     & mag (IMACS)  &  $L_z$                       & $0.013 \pm 0.001$        & $L_*$                   \\
	$g$            & $19.96 \pm 0.06$     & mag (3PI)    &  $L_{\rm H\alpha}$           & $15.5 \pm 1.4$           & $10^{38}$ erg s$^{-1}$  \\
	$g$            & $19.94 \pm 0.14$     & mag (LDSS)   &  $L_{\rm H\beta}$            & $6.8 \pm 1.8$            & $10^{38}$ erg s$^{-1}$  \\
	$r$            & $20.00 \pm 0.05$     & mag (3PI)    &  $L_{\rm [OII]_{3726,3729}}$ & $13.5 \pm 1.3$           & $10^{38}$ erg s$^{-1}$  \\
	$r$            & $20.04 \pm 0.02$     & mag (LDSS)   &  $L_{\rm [OIII]_{4959}}$     & $3.7 \pm 1.0$            & $10^{38}$ erg s$^{-1}$  \\
	$i$            & $19.96 \pm 0.09$     & mag (3PI)    &  $L_{\rm [OIII]_{5007}}$     & $12.2 \pm 2.6$           & $10^{38}$ erg s$^{-1}$  \\
	$i$            & $20.11 \pm 0.14$     & mag (LDSS)   &  $12 + \log($O/H$)$          & $7.80^{+0.15}_{-0.12}$   &                         \\
	$z$            & $19.98 \pm 0.12$     & mag (3PI)    &  $Z_{23}$                    & $0.14^{+0.05}_{-0.03}$   & Z$_\odot$               \\
	$y$            & $20.12 \pm 0.34$     & mag (3PI)    &  ${\rm SFR}_1$               & $0.49^{+0.20}_{-0.13}$   & M$_\odot$ yr$^{-1}$     \\
	$M_u$          & $-17.21 \pm 0.15$    & mag          &  ${\rm SFR}_2$               & $0.08^{+0.39}_{-0.03}$   & M$_\odot$ yr$^{-1}$     \\
	$M_g$          & $-17.53 \pm 0.06$    & mag          &  $\log(M)$                   & $8.56^{+0.17}_{-0.37}$   & M$_\odot$               \\
	$M_r$          & $-17.45 \pm 0.02$    & mag          &  $Z$                         & $0.08^{+0.06}_{-0.04}$   & Z$_\odot$               \\
	$M_i$          & $-17.53 \pm 0.09$    & mag          &                              &                          &                         \\
		\enddata
	\tablecomments{List of measured and derived properties of the host galaxy of SN\,2016iet. $P_{\rm cc}$ is the probability of chance coincidence, $z$ is the redshift, $D$ is the offset of SN\,2016iet from the host center, $r_{50}$ is the half light Petrosian radius in $r$-band, $v_r$ is the relative radial velocity difference between SN\,2016iet and the host, $ugrizy$ are aperture magnitudes corrected for Galactic extinction, $M_{ugriz}$ are the host absolute magnitudes, $L_{ugriz}$ are the host luminosities relative to $L_*$ in each band \citep{montero09}, $L_{n}$ are the integrated luminosities of several key emission lines (H$\alpha$, H$\beta$, [\ion{O}{2}], [\ion{O}{3}]),$12 + \log($O/H$)$ and $Z_{23}$ are values for the host metallicity inferred from the R23 formalism. SFR1 and SFR2 are the earliest 30 and 100 Myr of star formation from {\tt prospector} models. The mass and metallicity of the galaxy $\log(M)$ and $Z$ are also inferred from the {\tt prospector} models.}
\end{deluxetable*}

\begin{figure}
	\begin{center}
		\includegraphics[width=0.9\columnwidth]{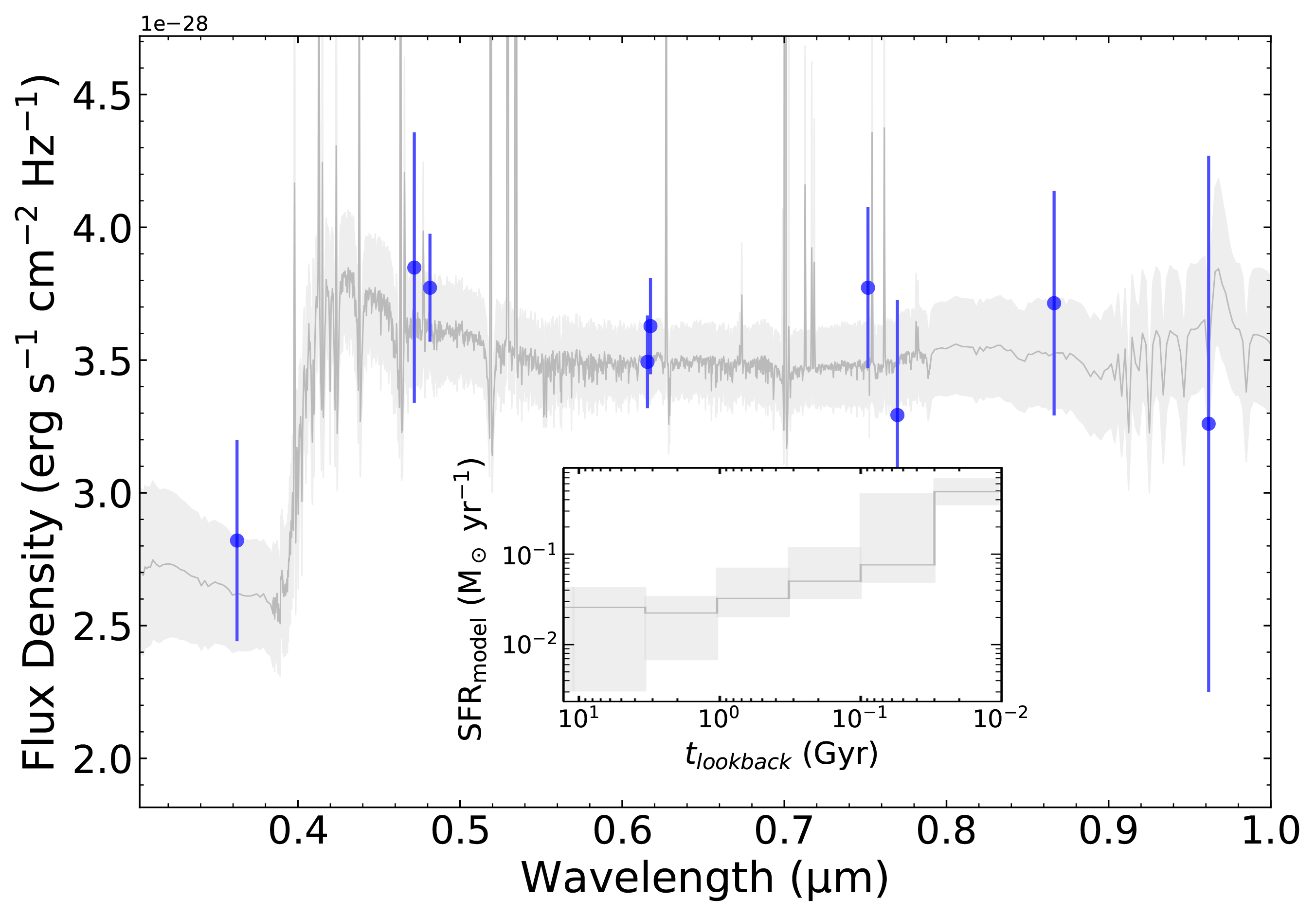}
		\caption{{Distribution of the most likely {\tt Prospector} models to the SED of the host galaxy of SN\,2016iet. The blue data points are the observed photometry listed in Table~\ref{tab:host}. The shaded grey region shows the $\pm1\sigma$ scatter in the posterior of {\tt Prospector} models. The insert shows the recovered star formation history. \label{fig:prospector}}}
	\end{center}
\end{figure}

To extract additional information about the host we model its SED using {\tt Prospector}, a software package designed to fit galaxy SEDs by taking into account dust attenuation and re-radiation, nebular emission, stellar metallicity, and a piecewise linear non-parametric star formation history \citep{leja17}. We model the SED with seven bins of star formation history, spaced evenly in logarithmic time from $t = 0$ to the age of the universe at $z = 0.0671$, allowing a two-component dust screen model with a flexible dust attenuation curve \citep{charlot00}. The resulting SED and {\tt Prospector} model are shown in Figure~\ref{fig:prospector}. We find a rising star formation history, where the most recent star formation rate is ${\rm SFR_{<30\,{\rm Myr}}} = 0.49^{+0.20}_{-0.13}$ M$_\odot$ yr$^{-1}$. We find a mass of $\log(M/$M$_\odot) = 8.56^{+0.17}_{-0.37}$, and a stellar metallicity of $Z = 0.08^{+0.06}_{-0.04}$, consistent with the gas-phase metallicity we inferred from emission line fluxes. We also fit a parameter permitting ionizing radiation to escape from the galaxy, instead of powering nebular emission. Inclusion of this parameter notably improves the quality of the fit to the photometry. We infer a high escape fraction $\gtrsim0.9$ from this method; we caution that this may be driven in part by deficiencies of the nebular emission model at low metallicities and high ionization states rather than true ionizing continuum leakage.

In our final spectrum at a phase of 772 days we detect unresolved H$\alpha$ emission at the location of SN\,2016iet (Figure~\ref{fig:halpha}), most likely originating from star formation. The H$\alpha$ luminosity corresponds to \mbox{${\rm SFR}\approx 3\times 10^{-4}$ M$_\odot$ yr$^{-1}$}. This is much smaller than for dwarf galaxies ($1.5$ M$_\odot$ yr$^{-1}$ for the SMC; \citealt{rezaeikh14}), but similiar to \ion{H}{2} regions in the SMC. For example, the NGC 346/N66 nebula, which has an \mbox{${\rm SFR}\approx 4\times 10^{-3}$ M$_\odot$ yr$^{-1}$} \citep{hony15}. Suggesting the progenitor of SN\,2016iet could have formed in-situ in an isolated \ion{H}{2} region. \cite{crowther13} finds that 7 out of 10 hydrogen poor SNe are associated with an \ion{H}{2} region. If an \ion{H}{2} region is present at the location of the SN, a very high mass progenitor ($>75$ M$_\odot$) is favored. Since for high-mass isolated clusters, gas has already been removed, we would not expect a SN coincident with the \ion{H}{2} region, unless this was from a very massive progenitor (i.e. have a lifetime shorter than the duty cycle of giant \ion{H}{2} regions.)

To summarize, SN\,2016iet is associated with a low metallicity dwarf galaxy, with an unusually large offset relative to other H/He-poor massive star explosions (Type~Ic SNe, SLSNe, and LGRBs). Since the range of possible progenitor core masses for SN\,2016iet is $55-120$M$_\odot$, which corresponds to ZAMS masses of $\sim 120-260$M$_\odot$, we conclude that the progenitor of SN\,2016iet must have formed in-situ. Stars of these masses have lifetimes of $\sim 3-4$ Myr (estimated from  stellar evolutionary tracks; \citealt{choi2016}).  If the large offset was due to ejection, this would require a velocity of \mbox{$\gtrsim 4000$ km s$^{-1}$}, much higher than even the fastest hypervelocity stars in the Milky Way (\mbox{$1200$ km s$^{-1}$}; \citealt{boubert18}). This therefore indicates that the progenitor of SN\,2016iet likely formed in-situ, in a dim satellite galaxy or an isolated star cluster. This is further evidenced by our detection of narrow H$\alpha$ at the location of the SN.

\section{SN\,2016iet as a PPISN or PISN} 
\label{sec:discussion} 

We infer that the progenitor of SN\,2016iet was a CO core with a mass near the end of its life of \mbox{$\gtrsim 55$ M$_\odot$} and potentially up to $120$ M$_\odot$. This is the range of core masses that are expected to produce PISNe or PPISNe \citep{woosley07}. We further find that the environment of SN\,2016iet has a metallicity of \mbox{$Z\approx 0.1$ Z$_\odot$}, in agreement with models for PISN and PPISN progenitors that suggest a threshold of $\lesssim 0.1$ Z$_\odot$ \citep{yusof13,owocki15,yoshida16}.  Within the context of our models, CSM interaction provides the most self-consistent explanation for the data; the radioactive model is not consistent with the spectra, while the engine models are not consistent with the notion that PISNe do not leave remnants.

Our minimal mass interaction model has an inferred CSM mass of about \mbox{$35$ M$_\odot$}, of which about \mbox{$33$ M$_\odot$} are located in a shell that spans about \mbox{$40-150$ AU} from the progenitor, and a few M$_\odot$ located at smaller radii (in the breakout region).  The ejecta mass is \mbox{$\approx 20$ M$_\odot$}, and the overall CO core mass is in the PPISN regime (Figure~\ref{fig:masses}).  The other interaction model variants have identical CSM masses and locations to the this model, but much larger ejecta masses of about \mbox{$45-85$ M$_\odot$} that place the progenitor in the PISN regime (Figure~\ref{fig:masses}).

Light curve models for PISNe from He cores, in which the emission is powered by radioactive decay and the diffusion of thermal energy (e.g., \citealt{kasen11}), do not match the data for SN\,2016iet.  \citet{kasen11} suggest that CSM interaction may provide an additional energy source, but they do not model this effect or make specific predictions for the structure of the CSM (e.g., mass, density, ejection timescale). To match the inferred properties of SN\,2016iet, stripped PISN progenitors would have to experience severe mass loss prior to the eventual explosion.     

Theoretical models of PPISNe have a wide range of predicted observational properties. \cite{woosley17} carried a detailed study of PPISNe from helium cores of $30-64$ M$_\odot$ and their pulsation histories. We compare SN\,2016iet to the PPISN model of a $62$ M$_\odot$ helium core. In this model $28$ M$_\odot$ of material are ejected in a pulsational mass loss event about 2700 years prior to core collapse, leaving behind a $34$ M$_\odot$ remnant that may experience the mass loss of an additional few M$_\odot$ in the final years before explosion, and then most likely collapses to a black hole.  While the large mass ejection and remaining core mass are comparable to what we infer for SN\,2016iet, the mass ejection timescale is off by nearly three orders of magnitude.  Moreover, our inference of a large ejecta mass for SN\,2016iet would require the core-collapse explosion to be successful.

Thus, while at face value SN\,2016iet matches the predictions for PPISN/PISN explosions in terms of the inferred progenitor mass and metallicity, in detail, it appears to challenge these models by requiring that about half of the progenitor's mass be lost in the final years before explosion. This suggests that either SN\,2016iet was formed by a completely different mechanism, or that PPISNe/PISNe progenitors may undergo different behavior than suggested by current theoretical models. 

\section{Conclusions} 
\label{sec:conclusion}

We presented detailed optical follow-up of SN\,2016iet, including the determination of its redshift and identification of its host galaxy. We classify this event as a unique Type~I SN due to the lack of H and He. The combination of its unusual light curve, spectra, and environment is unprecedented; the only potential analogue in the existing literature is SN\,2010mb, but even it does not share all of the same properties (e.g., luminosity, duration, and low metallicity host).  The key observational properties of SN\,2016iet are:

\begin{itemize}
	\item The light curve exhibits a double-peaked structure with nearly equal brightness peaks of about $-19$ mag separated by about 100 rest-frame days.
    \item The subsequent decline in brightness is slow, $\approx 5$ mag in $\approx 650$ rest-frame days, with the SN remaining detectable at the present.
	\item The spectra are dominated by intermediate width ($\approx 3400$ km s$^{-1}$) emission lines of calcium and oxygen, superposed on a strong blue continuum in the first year, and lack hydrogen or helium.
    \item The ratio of [\ion{Ca}{2}] to \ion{O}{1} of about 4 in the late (nebular) spectra is much larger than for normal stripped core-collapse SNe and SLSNe.  It is reminiscent of Ca-rich transients, although the duration and luminosity of Ca-rich transients are far lower than those of SN\,2016iet.
    \item The forbidden oxygen lines are redshifted by \mbox{$\approx 2800$ km s$^{-1}$} with respect to the permitted oxygen lines and permitted and forbidden calcium lines. The observed redshift may be indicative of a light echo from a circumstellar shell not yet reached by the ejecta.
	\item The SN is located at a large offset ($16.5$ kpc, $4.3$ effective radii) from a low metallicity dwarf galaxy. This offset is substantially larger than for other types of stripped envelope explosions (Type~Ib/c SNe, SLSNe, LGRBs).
	\item The host galaxy has a low metallicity of \mbox{$Z\approx 0.1$ Z$_\odot$}, unusually low even in comparison to SLSN and LGRB host galaxies.	
	\item The detection of a weak and narrow H$\alpha$ line at late time (phase of 772 days) coincident with the SN location indicates underlying star formation activity, either from a dwarf galaxy with $L\lesssim 0.003$ $L_*$, or an isolated star forming region.
\end{itemize}

We explore several models for the energy source powering SN\,2016iet -- radioactive decay, central engine, and CSM interaction -- and find that regardless of the model, the inferred mass of the progenitor (CO core) prior to explosion is in the range of $\approx 55-120$ M$_\odot$.  This suggests a ZAMS mass of $\approx 120-260$ M$_\odot$, which along with the low metallicity, places the progenitor in the regime of PPISN/PISN explosions. {\it This is the first PPISNe/PISN candidate with an inferred progenitor mass and metallicity in the range predicted by theoretical models.}  

In detail, the most consistent model for the light curve and spectra is CSM interaction, in which the CSM has a mass of $\approx 35$ M$_\odot$ and is located over a range of about $40-150$ AU of the progenitor.  This indicates that the CSM was ejected within a decade before the explosion at a rate of $\approx 7$ M$_\odot$ yr$^{-1}$.  In the model of breakout+interaction, an additional few M$_\odot$ of material are required in even closer vicinity to the progenitor. 

While the mass of the progenitor and its inferred metallicity are overall consistent with the predictions of PPISN/PISN models, we find that in detail, the inferred properties are challenging to explain in the framework of existing predictions.  In particular, PPISN models in the range of masses inferred here can indeed eject about half of their CO core mass, but are predicted to do so thousands of years before explosion \citep{woosley17}, not in less than a decade, as required for SN\,2016iet.  Additional mass loss may occur in the final years before the final core collapse, but at a level of only a few M$_\odot$ \citep{woosley17,aguilera18}, an order of magnitude too low for SN\,2016iet.  Finally, \citet{woosley17} also argue that the remaining cores of \mbox{$\sim 35-45$ M$_\odot$} have a large binding energy that may make them difficult to explode, and lead instead of massive black holes; this is again in tension with the inferred ejecta mass of \mbox{$\sim 20-85$ M$_\odot$} for SN\,2016iet. In the context of PISN models, CSM interaction has been mentioned as a potential energy source for explosions of He cores, but this has not been explored in detail (e.g., \citealt{kasen11}).  

We end with the following cautionary note.  SN\,2016iet may indicate that existing predictions for PPISN/PISN explosions are flawed. Namely, either PPISN/PISN progenitors can eject about half of their mass in the final years before explosion (rather than several thousand years before explosion), or PISNe can actually leave behind compact remnants in the form of a millisecond magnetar or an accreting compact object (as required in the context of the engine models for SN\,2016iet). Alternatively, SN\,2016iet could instead be powered by a completely different mechanism not explored in this work, such as a binary merger involving at least one CO core.

\acknowledgments
We thank D.~Aguilera-Dena, J.~Guillochon, D.~Kasen, and S.~Woosley for useful discussions, and Y.~Beletsky for carrying out some of the Magellan observations. The Berger Time-Domain Group at Harvard is supported in part by NSF under grant AST-1714498 and by NASA under grant NNX15AE50G. S.~Gomez is partly supported by an NSF Graduate Research Fellowship. M.~Nicholl is supported by a Royal Astronomical Society Research Fellowship.  This work is based in part on observations obtained at the MDM Observatory, operated by Dartmouth College, Columbia University, Ohio State University, Ohio University, and the University of Michigan. We acknowledge ESA Gaia, DPAC and the Photometric Science Alerts Team (http://gsaweb.ast.cam.ac.uk/alerts). This paper includes data gathered with the 6.5 meter Magellan Telescopes located at Las Campanas Observatory, Chile. Observations reported here were obtained at the MMT Observatory, a joint facility of the University of Arizona and the Smithsonian Institution. This research has made use of NASA’s Astrophysics Data System. This research has made use of the SIMBAD database, operated at CDS, Strasbourg, France. Based on observations obtained at the Gemini Observatory (program GN-2019A-DD-103), processed using the Gemini IRAF package, which is operated by the Association of Universities for Research in Astronomy, Inc., under a cooperative agreement with the NSF on behalf of the Gemini partnership: the National Science Foundation (United States), National Research Council (Canada), CONICYT (Chile), Ministerio de Ciencia, Tecnolog\'{i}a e Innovaci\'{o}n Productiva (Argentina), Minist\'{e}rio da Ci\^{e}ncia, Tecnologia e Inova\c{c}\~{a}o (Brazil), and Korea Astronomy and Space Science Institute (Republic of Korea). 

\software{Astropy\citep{astropy18}, MOSFiT\citep{guillochon18}, Pyraf\citep{science12}, SAOImage DS9 \citep{sao00}, emcee\citep{foreman13}, corner \citep{foreman16}, Superbol \citep{nicholl18b}, Matplotlib\citep{hunter2007}, SciPy\citep{jones01}, numpy\citep{oliphant06}, extinction(\url{https://github.com/kbarbary/extinction}), PYPHOT(\url{https://github.com/mfouesneau/pyphot}}).



\appendix
\setcounter{figure}{0}
\setcounter{table}{0}

\section{Data Tables}\label{sec:app_bg_treatment}

\startlongtable
\begin{deluxetable*}{cccccccc}
	\tabletypesize{\tiny}
	\tablecaption{Optical Photometry \label{tab:photometry}}
	\tablewidth{0pt}
	\tablehead{
		\colhead{UT Date} &		\colhead{MJD} &		\colhead{$g$ (err)} &	\colhead{$r$ (err)} & \colhead{$i$ (err)} & \colhead{$C$ (err) }& \colhead{\textit{G} (err)} & Telescope/Instrument  \\
		(year-month-day) & 							  &       \colhead{(mag)}    &   \colhead{(mag)} 	 & \colhead{(mag)} 	  & \colhead{(mag)} 	  & \colhead{(mag)}    & 
	}
	\startdata
	2016-05-19 & 57527.4 & \nodata     & \nodata     &$> 19.7 $    & \nodata     &  \nodata  & PSST \\
	2016-07-11 & 57580.8 & \nodata     & \nodata     & \nodata     & \nodata     & $> 20.7$ & Gaia \\
	2016-11-14 & 57706.6 & \nodata     & \nodata     & \nodata     & \nodata     & 18.61(0.01) & Gaia \\
	2016-11-14 & 57706.8 & \nodata     & \nodata     & \nodata     & \nodata     & 18.59(0.01) & Gaia \\
	2016-12-10 & 57732.0 & \nodata     & \nodata     & \nodata     & \nodata     & 18.20(0.01) & Gaia \\
	2016-12-10 & 57732.1 & \nodata     & \nodata     & \nodata     & \nodata     & 18.23(0.01) & Gaia \\
	2017-01-05 & 57758.3 & \nodata     & \nodata     & \nodata     & \nodata     & 18.73(0.01) & Gaia \\
	2017-02-01 & 57785.5 & \nodata     & \nodata     & \nodata     & 19.03(0.12) & \nodata     & MLS \\
	2017-02-01 & 57785.5 & \nodata     & \nodata     & \nodata     & 18.94(0.11) & \nodata     & MLS \\
	2017-02-01 & 57785.5 & \nodata     & \nodata     & \nodata     & 18.88(0.11) & \nodata     & MLS \\
	2017-02-01 & 57785.5 & \nodata     & \nodata     & \nodata     & 18.85(0.11) & \nodata     & MLS \\
	2017-02-23 & 57807.5 & \nodata     & \nodata     & \nodata     & 19.04(0.09) & \nodata     & CRTS \\
	2017-02-23 & 57807.5 & \nodata     & \nodata     & \nodata     & 19.01(0.09) & \nodata     & CRTS \\
	2017-02-23 & 57807.5 & \nodata     & \nodata     & \nodata     & 18.97(0.08) & \nodata     & CRTS \\
	2017-02-23 & 57807.5 & \nodata     & \nodata     & \nodata     & 19.02(0.09) & \nodata     & CRTS \\
	2017-03-06 & 57818.4 & \nodata     & \nodata     & 18.75(0.02) & \nodata     & \nodata     & PSST \\
	2017-03-06 & 57818.4 & \nodata     & \nodata     & 18.75(0.02) & \nodata     & \nodata     & PSST \\
	2017-03-06 & 57818.5 & \nodata     & \nodata     & 18.79(0.03) & \nodata     & \nodata     & PSST \\
	2017-03-06 & 57818.5 & \nodata     & \nodata     & 18.77(0.02) & \nodata     & \nodata     & PSST \\
	2017-03-06 & 57818.5 & \nodata     & \nodata     & 18.77(0.03) & \nodata     & \nodata     & PSST \\
	2017-03-16 & 57828.5 & \nodata     & \nodata     & \nodata     & 18.57(0.08) & \nodata     & CRTS \\
	2017-03-16 & 57828.5 & \nodata     & \nodata     & \nodata     & 18.71(0.08) & \nodata     & CRTS \\
	2017-03-16 & 57828.5 & \nodata     & \nodata     & \nodata     & 18.75(0.08) & \nodata     & CRTS \\
	2017-03-16 & 57828.5 & \nodata     & \nodata     & \nodata     & 18.70(0.08) & \nodata     & CRTS \\
	2017-03-16 & 57828.5 & \nodata     & \nodata     & 18.78(0.05) & \nodata     & \nodata     & PSST \\
	2017-03-16 & 57828.5 & \nodata     & \nodata     & 18.76(0.04) & \nodata     & \nodata     & PSST \\
	2017-03-16 & 57828.6 & \nodata     & \nodata     & 18.80(0.06) & \nodata     & \nodata     & PSST \\
	2017-03-16 & 57828.6 & \nodata     & \nodata     & 18.80(0.06) & \nodata     & \nodata     & PSST \\
	2017-03-25 & 57837.5 & \nodata     & \nodata     & \nodata     & 18.73(0.08) & \nodata     & CRTS \\
	2017-03-25 & 57837.5 & \nodata     & \nodata     & \nodata     & 18.72(0.08) & \nodata     & CRTS \\
	2017-03-25 & 57837.5 & \nodata     & \nodata     & \nodata     & 18.64(0.08) & \nodata     & CRTS \\
	2017-03-25 & 57837.5 & \nodata     & \nodata     & \nodata     & 18.66(0.08) & \nodata     & CRTS \\
	2017-03-31 & 57843.5 & \nodata     & \nodata     & \nodata     & 18.75(0.08) & \nodata     & CRTS \\
	2017-03-31 & 57843.5 & \nodata     & \nodata     & \nodata     & 18.72(0.08) & \nodata     & CRTS \\
	2017-03-31 & 57843.5 & \nodata     & \nodata     & \nodata     & 18.67(0.08) & \nodata     & CRTS \\
	2017-03-31 & 57843.5 & \nodata     & \nodata     & \nodata     & 18.72(0.08) & \nodata     & CRTS \\
	2017-04-15 & 57858.5 & \nodata     & \nodata     & 18.91(0.06) & \nodata     & \nodata     & PSST \\
	2017-04-15 & 57858.5 & \nodata     & \nodata     & 18.84(0.06) & \nodata     & \nodata     & PSST \\
	2017-04-15 & 57858.5 & \nodata     & \nodata     & 18.86(0.07) & \nodata     & \nodata     & PSST \\
	2017-04-15 & 57858.5 & \nodata     & \nodata     & 18.86(0.06) & \nodata     & \nodata     & PSST \\
	2017-04-15 & 57858.5 & \nodata     & \nodata     & \nodata     & 19.09(0.13) & \nodata     & MLS \\
	2017-04-15 & 57858.5 & \nodata     & \nodata     & \nodata     & 18.78(0.11) & \nodata     & MLS \\
	2017-04-15 & 57858.5 & \nodata     & \nodata     & \nodata     & 18.94(0.12) & \nodata     & MLS \\
	2017-04-15 & 57858.5 & \nodata     & \nodata     & \nodata     & 18.96(0.12) & \nodata     & MLS \\
	2017-04-21 & 57864.5 & \nodata     & \nodata     & \nodata     & 18.88(0.08) & \nodata     & CRTS \\
	2017-04-21 & 57864.5 & \nodata     & \nodata     & \nodata     & 18.94(0.08) & \nodata     & CRTS \\
	2017-04-21 & 57864.5 & \nodata     & \nodata     & \nodata     & 18.86(0.08) & \nodata     & CRTS \\
	2017-04-21 & 57864.5 & \nodata     & \nodata     & \nodata     & 18.87(0.08) & \nodata     & CRTS \\
	2017-05-21 & 57894.5 & \nodata     & \nodata     & \nodata     & 19.01(0.09) & \nodata     & CRTS \\
	2017-05-21 & 57894.5 & \nodata     & \nodata     & \nodata     & 19.21(0.09) & \nodata     & CRTS \\
	2017-05-21 & 57894.5 & \nodata     & \nodata     & \nodata     & 19.12(0.09) & \nodata     & CRTS \\
	2017-05-21 & 57894.5 & \nodata     & \nodata     & \nodata     & 19.10(0.09) & \nodata     & CRTS \\
	2017-05-23 & 57896.3 & 19.14(0.10) & 19.07(0.09) & 19.23(0.13) & \nodata     & \nodata     & FLWO48 \\
	2017-05-24 & 57897.3 & 19.22(0.10) & 19.10(0.10) & 19.10(0.13) & \nodata     & \nodata     & FLWO48 \\
	2017-05-28 & 57901.3 & 19.23(0.09) & 19.17(0.09) & 19.21(0.15) & \nodata     & \nodata     & FLWO48 \\
	2017-05-29 & 57902.3 & 19.35(0.10) & 19.28(0.13) & 19.40(0.15) & \nodata     & \nodata     & FLWO48 \\
	2017-06-06 & 57910.2 & 19.29(0.11) & 19.24(0.14) & 19.29(0.12) & \nodata     & \nodata     & FLWO48 \\
	2017-06-06 & 57910.3 & \nodata     & \nodata     & 19.37(0.15) & \nodata     & \nodata     & PSST \\
	2017-06-06 & 57910.3 & \nodata     & \nodata     & 19.38(0.15) & \nodata     & \nodata     & PSST \\
	2017-06-06 & 57910.3 & \nodata     & \nodata     & 19.28(0.15) & \nodata     & \nodata     & PSST \\
	2017-06-11 & 57915.2 & 19.31(0.06) & 19.26(0.09) & 19.27(0.13) & \nodata     & \nodata     & FLWO48 \\
	2017-06-12 & 57916.2 & 19.35(0.05) & 19.27(0.10) & 19.22(0.11) & \nodata     & \nodata     & FLWO48 \\
	2017-06-13 & 57917.2 & 19.34(0.11) & 19.32(0.11) & 19.29(0.14) & \nodata     & \nodata     & FLWO48 \\
	2017-06-16 & 57920.2 & 19.53(0.12) & 19.40(0.04) & 19.31(0.12) & \nodata     & \nodata     & MDM \\
	2017-06-18 & 57922.3 & \nodata     & 19.40(0.09) & 19.31(0.13) & \nodata     & \nodata     & MDM \\
	2017-06-30 & 57934.1 & 19.61(0.14) & \nodata     & \nodata     & \nodata     & \nodata     & FLWO48 \\
	2017-06-30 & 57934.2 & 19.54(0.09) & 19.41(0.11) & 19.06(0.22) & \nodata     & \nodata     & FLWO48 \\
	2017-06-30 & 57934.2 & \nodata     & 19.44(0.12) & 19.30(0.14) & \nodata     & \nodata     & FLWO48 \\
	2017-07-02 & 57936.1 & 19.39(0.11) & \nodata     & \nodata     & \nodata     & \nodata     & FLWO48 \\
	2017-07-02 & 57936.2 & 19.51(0.10) & 19.44(0.11) & 19.41(0.11) & \nodata     & \nodata     & FLWO48 \\
	2017-07-02 & 57936.2 & \nodata     & 19.51(0.12) & 19.35(0.12) & \nodata     & \nodata     & FLWO48 \\
	2017-07-02 & 57936.5 & \nodata     & \nodata     & \nodata     & 19.32(0.09) & \nodata     & CRTS \\
	2017-07-02 & 57936.5 & \nodata     & \nodata     & \nodata     & 19.32(0.09) & \nodata     & CRTS \\
	2017-07-02 & 57936.5 & \nodata     & \nodata     & \nodata     & 19.38(0.09) & \nodata     & CRTS \\
	2017-07-02 & 57936.5 & \nodata     & \nodata     & \nodata     & 19.24(0.09) & \nodata     & CRTS \\
	2017-07-08 & 57942.2 & 19.60(0.20) & 19.20(0.11) & \nodata     & \nodata     & \nodata     & FLWO48 \\
	2017-07-08 & 57942.2 & 19.47(0.11) & 19.18(0.16) & \nodata     & \nodata     & \nodata     & FLWO48 \\
	2017-07-22 & 57956.9 & \nodata     & \nodata     & \nodata     & \nodata     & 19.34(0.01) & Gaia \\
	2017-07-22 & 57956.9 & \nodata     & \nodata     & \nodata     & \nodata     & 19.40(0.01) & Gaia \\
	2017-11-15 & 58072.5 & 21.12(0.10) & 20.92(0.15) & 20.67(0.16) & \nodata     & \nodata     & FLWO48 \\
	2017-11-15 & 58072.5 & 20.98(0.12) & 20.85(0.10) & 20.45(0.15) & \nodata     & \nodata     & FLWO48 \\
	2017-11-19 & 58076.5 & 21.24(0.12) & 20.98(0.13) & 20.70(0.18) & \nodata     & \nodata     & FLWO48 \\
	2017-11-19 & 58076.5 & 21.25(0.12) & 21.03(0.09) & 20.74(0.20) & \nodata     & \nodata     & FLWO48 \\
	2017-11-19 & 58076.5 & 21.27(0.12) & 20.95(0.15) & 20.49(0.19) & \nodata     & \nodata     & FLWO48 \\
	2018-01-31 & 58149.5 & \nodata     & 22.14(0.22) & \nodata     & \nodata     & \nodata     & MMTCam \\
	2018-01-31 & 58149.5 & 22.23(0.10) & \nodata     & \nodata     & \nodata     & \nodata     & FLWO48 \\
	2018-01-31 & 58149.6 & \nodata     & \nodata     & 21.41(0.19) & \nodata     & \nodata     & MMTCam \\
	2018-02-05 & 58154.5 & 22.36(0.18) & 21.77(0.27) & 21.62(0.17) & \nodata     & \nodata     & MMTCam \\
	2018-02-25 & 58174.4 & 22.43(0.11) & 22.11(0.09) & 21.50(0.10) & \nodata     & \nodata     & MMTCam \\
	2018-02-26 & 58175.4 & 22.31(0.20) & 22.06(0.12) & \nodata     & \nodata     & \nodata     & MMTCam \\
	2018-03-16 & 58193.4 & \nodata     & \nodata     & 21.61(0.11) & \nodata     & \nodata     & MMTCam \\
	2018-03-18 & 58195.2 & 22.38(0.25) & \nodata     & \nodata     & \nodata     & \nodata     & LDSS \\
	2018-03-18 & 58195.3 & \nodata     & 22.08(0.09) & 21.57(0.15) & \nodata     & \nodata     & LDSS \\
	2018-04-05 & 58213.4 & 22.33(0.31) & 22.27(0.32) & \nodata     & \nodata     & \nodata     & FLWO48 \\
	2018-04-05 & 58213.5 & \nodata     & \nodata     & 21.64(0.32) & \nodata     & \nodata     & FLWO48 \\
	2018-05-17 & 58255.3 & \nodata     & 21.98(0.12) & 21.56(0.11) & \nodata     & \nodata     & MMTCam \\
	2018-07-09 & 58309.0 & \nodata     & \nodata     & 21.95(0.28) & \nodata     & \nodata     & LDSS \\
	2018-07-09 & 58309.0 & \nodata     & \nodata     & 21.95(0.28) & \nodata     & \nodata     & LDSS \\
	2018-07-09 & 58309.0 & \nodata     & \nodata     & 21.95(0.28) & \nodata     & \nodata     & LDSS \\
    2018-12-09 & 58461.5 & \nodata     &  22.73 (0.44)   & \nodata   & \nodata   & \nodata   & MDM \\
	2019-01-31 & 58514.3 & \nodata     & 23.43 (0.08)    & \nodata  & \nodata     & \nodata     & LDSS \\
    2019-01-31 & 58514.3 & \nodata     & \nodata     & 23.52 (0.08) & \nodata     & \nodata     & LDSS \\
	\enddata          
	\tablecomments{Optical photometry of SN\,2016iet, not corrected for galactic extinction. Optical $gri$ magnitudes are in the AB system, $C$ magnitudes are calibrated to Vega magnitudes, and \textit{G} magnitudes come from the Gaia Phtometric Alerts Index. The photometric error bars are shown in parenthesis. The ``$>$" represents a $3\sigma$ upper limit. }
\end{deluxetable*}

\begin{deluxetable*}{ccccccccc}
	\tabletypesize{\tiny}
	\tablecaption{Optical Spectroscopy of SN\,2016iet \label{tab:spectroscopy}}
	\tablewidth{0pt}
	\tablehead{
		\colhead{UT Date} & \colhead{MJD} & \colhead{Phase\tablenotemark{a}} & \colhead{Exposure Time} & \colhead{Airmass} & Wavelength Range & \colhead{Telescope + Instrument} & \colhead{Grating} & \colhead{Slit width} \\
		& 			  & \colhead{(d)}   & \colhead{(s)}			  &					  &		(\AA)			&				 & (lines/mm)		 & (arcsec)						 
	}
	\startdata
    2017 Apr 30  & 57873.00  & +132 & 1200    & 1.79    & 4160-9340  & Baade + IMACS      & 300     & 0.9"    \\
    2017 May 23 & 57896.33  & +153 & 1200    & 1.481   & 3310-8480  & MMT + Blue Channel & 300     & 1"      \\
    2017 Jun 17  & 57921.00  & +177  &  3600 & 1.70 & 3920-9050  & MDM + OSMOS        & 1600 & 1" \\
    2017 Jun 29 & 57933.22  & +187 & 1200    & 1.51    & 3320-8560  & MMT + Blue Channel & 300     & 1"      \\
    2017 Nov 26 & 58083.51  & +328 & 1200    & 1.53    & 3320-8590  & MMT + Blue Channel & 300     & 1"      \\
    2018 Jan 27  & 58145.44  & +386 & 1800    & 1.07    & 3290-8550  & MMT + Blue Channel & 300     & 1"      \\
    2018 Feb 9    & 58158.50  & +399 & 1200    & 1.05    & 3820-9200  & MMT + Binospec     & 270     & 1"      \\
    2018 Mar 17   & 58194.24  & +433 & 1800    & 1.79    & 3900-10600 & Clay + LDSS3c      & 400     & 1"      \\
    2019 Mar 4   & 58556.58  & +772 & 7200    & 1.05    & 5300-10200 & Gemini + GMOS      & 400     & 1"      \\
   	\enddata
	\tablenotetext{a}{Rest-frame days since peak bolometric brightness.}
\end{deluxetable*}


\begin{thebibliography}{}

\bibitem[Aguilera-Dena et al.(2018)]{aguilera18} Aguilera-Dena, D.~R., Langer, N., Moriya, T.~J., \& Schootemeijer, A.\ 2018, \apj, 858, 115 
\bibitem[Arcavi et al.(2014)]{arcavi14} Arcavi, I., Gal-Yam, A., Sullivan, M., et al.\ 2014, \apj, 793, 38 
\bibitem[Arcavi et al.(2017)]{arcavi17} Arcavi, I., Howell, D.~A., Kasen, D., et al.\ 2017, \nat, 551, 210 
\bibitem[Asplund et al.(2009)]{asplund09} Asplund, M., Grevesse, N., Sauval, A.~J., \& Scott, P.\ 2009, \araa, 47, 481 
\bibitem[Astropy Collaboration et al.(2018)]{astropy18} Astropy Collaboration, Price-Whelan, A.~M., Sip{\H o}cz, B.~M., et al.\ 2018, \aj, 156, 123 
\bibitem[Barkat et al.(1967)]{barkat67} Barkat, Z., Rakavy, G., \& Sack, N.\ 1967, Physical Review Letters, 18, 379 
\bibitem[Ben-Ami et al.(2014)]{benami14} Ben-Ami, S., Gal-Yam, A., Mazzali, P.~A., et al.\ 2014, \apj, 785, 37 
\bibitem[Blanchard et al.(2016)]{blanchard16} Blanchard, P.~K., Berger, E., \& Fong, W.-f.\ 2016, \apj, 817, 144 
\bibitem[Blanchard et al.(2018)]{blanchard18} Blanchard, P.~K., Nicholl, M., Berger, E., et al.\ 2018, \apj, 865, 9 
\bibitem[Boubert et al.(2018)]{boubert18} Boubert, D., Guillochon, J., Hawkins, K., et al.\ 2018, \mnras, 479, 2789 
\bibitem[Burrows et al.(2005)]{burrows05} Burrows, D.~N., Hill, J.~E., Nousek, J.~A., et al.\ 2005, \ssr, 120, 165
\bibitem[Charlot \& Fall(2000)]{charlot00} Charlot, S., \& Fall, S.~M.\ 2000, \apj, 539, 718 
\bibitem[Chatzopoulos et al.(2013)]{chatzopoulos13} Chatzopoulos, E., Wheeler, J.~C., Vinko, J., Horvath, Z.~L., \& Nagy, A.\ 2013, \apj, 773, 76 
\bibitem[Chevalier \& Irwin(2011)]{chevalier11} Chevalier, R.~A., \& Irwin, C.~M.\ 2011, \apjl, 729, L6 
\bibitem[Chomiuk et al.(2011)]{chomiuk11} Chomiuk, L., Chornock, R., Soderberg, A.~M., et al.\ 2011, \apj, 743, 114 
\bibitem[Chornock et al.(2014)]{chornock14} Chornock, R., Berger, E., Gezari, S., et al.\ 2014, \apj, 780, 44 
\bibitem[Choi et al.(2016)]{choi2016} Choi, J., Dotter, A., Conroy, C., et al.\ 2016, \apj, 823, 102 
\bibitem[Crowther(2013)]{crowther13} Crowther, P.~A.\ 2013, \mnras, 428, 1927.
\bibitem[De et al.(2018)]{de2018} De, K., Kasliwal, M.~M., Cantwell, T., et al.\ 2018, \apj, 866, 72 
\bibitem[Dessart et al.(2013)]{dessart13} Dessart, L., Waldman, R., Livne, E., Hillier, D.~J., \& Blondin, S.\ 2013, \mnras, 428, 3227 
\bibitem[Dexter \& Kasen(2013)]{dexter13} Dexter, J., \& Kasen, D.\ 2013, \apj, 772, 30 
\bibitem[Drake et al.(2009)]{drake09} Drake, A.~J., Djorgovski, S.~G., Mahabal, A., et al.\ 2009, \apj, 696, 870 
\bibitem[Dressler et al.(2011)]{dressler11} Dressler, A., Bigelow, B., Hare, T., et al.\ 2011, \pasp, 123, 288 
\bibitem[Drout et al.(2014)]{drout14} Drout, M.~R., Chornock, R., Soderberg, A.~M., et al.\ 2014, \apj, 794, 23 
\bibitem[Fabricant et al.(2003)]{fabricant03} Fabricant, D.~G., Epps, H.~W., Brown, W.~L., Fata, R.~G., \& Mueller, M.\ 2003, \procspie, 4841, 1134 
\bibitem[Foreman-Mackey et al.(2013)]{foreman13} Foreman-Mackey, D., Hogg, D.~W., Lang, D., \& Goodman, J.\ 2013, \pasp, 125, 306 
\bibitem[Foreman-Mackey(2016)]{foreman16} Foreman-Mackey, D.\ 2016, The Journal of Open Source Software, 1,  
\bibitem[Gal-Yam et al.(2009)]{galyam09} Gal-Yam, A., Mazzali, P., Ofek, E.~O., et al.\ 2009, \nat, 462, 624 
\bibitem[Galama et al.(1998)]{galama98} Galama, T.~J., Vreeswijk, P.~M., van Paradijs, J., et al.\ 1998, \nat, 395, 670 
\bibitem[Gelman \& Rubin(1992)]{gelman92} Gelman, A., \& Rubin, D.~B.\ 1992, Statistical Science, 7, 457 
\bibitem[Gezari et al.(2009)]{gezari09} Gezari, S., Heckman, T., Cenko, S.~B., et al.\ 2009, \apj, 698, 1367 
\bibitem[Guillochon et al.(2017)]{guillochon17} Guillochon, J., Parrent, J., Kelley, L.~Z., \& Margutti, R.\ 2017, \apj, 835, 64 
\bibitem[Guillochon et al.(2018)]{guillochon18} Guillochon, J., Nicholl, M., Villar, V.~A., et al.\ 2018, \apjs, 236, 6 
\bibitem[Heger et al.(2002)]{heger02} Heger, A., Woosley, S., Baraffe, I., \& Abel, T.\ 2002, Lighthouses of the Universe: The Most Luminous Celestial Objects and Their Use for Cosmology, 369 
\bibitem[Hinshaw et al.(2013)]{hinshaw13} Hinshaw, G., Larson, D., Komatsu, E., et al.\ 2013, \apjs, 208, 19 
\bibitem[Hony et al.(2015)]{hony15} Hony, S., Gouliermis, D.~A., Galliano, F., et al.\ 2015, \mnras, 448, 1847 
\bibitem[Hook et al.(2004)]{hook04} Hook, I.~M., J{\o}rgensen, I., Allington-Smith, J.~R., et al.\ 2004, \pasp, 116, 425 
\bibitem[Huber et al.(2015)]{huber15} Huber, M., Chambers, K.~C., Flewelling, H., et al.\ 2015, The Astronomer's Telegram, 7153
\bibitem[Hunter (2007)]{hunter2007} Hunter, J. D.\ 2007, Computing in Science \& Engineering, 9, 90
\bibitem[Inserra et al.(2013)]{inserra13} Inserra, C., Smartt, S.~J., Jerkstrand, A., et al.\ 2013, \apj, 770, 128 
\bibitem[Jerkstrand et al.(2016)]{jerkstrand16} Jerkstrand, A., Smartt, S.~J., \& Heger, A.\ 2016, \mnras, 455, 3207 
\bibitem[Jerkstrand(2017)]{jerkstrand17} Jerkstrand, A.\ 2017, Handbook of Supernovae, 795 
\bibitem[Jones et al.(2001)]{jones01} Jones, E., Oliphant, T., Peterson, P., et al.\ 2001, SciPy
\bibitem[Karamehmetoglu et al.(2017)]{karamehmetoglu17} Karamehmetoglu, E., Taddia, F., Sollerman, J., et al.\ 2017, \aap, 602, A93 
\bibitem[Kasliwal et al.(2012)]{kasliwal12} Kasliwal, M.~M., Kulkarni, S.~R., Gal-Yam, A., et al.\ 2012, \apj, 755, 161 
\bibitem[Kasen \& Bildsten(2010)]{kasen10} Kasen, D., \& Bildsten, L.\ 2010, \apj, 717, 245 
\bibitem[Kasen et al.(2011)]{kasen11} Kasen, D., Woosley, S.~E., \& Heger, A.\ 2011, \apj, 734, 102 
\bibitem[Kelly \& Kirshner(2012)]{kelly12} Kelly, P.~L., \& Kirshner, R.~P.\ 2012, \apj, 759, 107 
\bibitem[Kennicutt(1998)]{kennicutt98} Kennicutt, R.~C., Jr.\ 1998, \araa, 36, 189 
\bibitem[Kobulnicky et al.(1999)]{kobulnicky99} Kobulnicky, H.~A., Kennicutt, R.~C., Jr., \& Pizagno, J.~L.\ 1999, \apj, 514, 544 
\bibitem[Leja et al.(2017)]{leja17} Leja, J., Johnson, B.~D., Conroy, C., van Dokkum, P.~G., \& Byler, N.\ 2017, \apj, 837, 170 
\bibitem[Leloudas et al.(2012)]{leloudas12} Leloudas, G., Chatzopoulos, E., Dilday, B., et al.\ 2012, \aap, 541, A129
\bibitem[Lunnan et al.(2014)]{lunnan14} Lunnan, R., Chornock, R., Berger, E., et al.\ 2014, \apj, 787, 138 
\bibitem[Lunnan et al.(2015)]{lunnan15} Lunnan, R., Chornock, R., Berger, E., et al.\ 2015, \apj, 804, 90 
\bibitem[Lunnan et al.(2016)]{lunnan16} Lunnan, R., Chornock, R., Berger, E., et al.\ 2016, \apj, 831, 144 
\bibitem[Lunnan et al.(2017)]{lunnan17} Lunnan, R., Kasliwal, M.~M., Cao, Y., et al.\ 2017, \apj, 836, 60 
\bibitem[Lunnan et al.(2018)]{lunnan18} Lunnan, R., Fransson, C., Vreesqijk, P.~M., et al.\ 2018, Nature Astronomy, 2, 887 
\bibitem[Martini et al.(2011)]{martini11} Martini, P., Stoll, R., Derwent, M.~A., et al.\ 2011, Publications of the Astronomical Society of the Pacific, 123, 187
\bibitem[Mazzali et al.(2019)]{mazzali19} Mazzali, P.~A., Moriya, T.~J., Tanaka, M., \& Woosley, S.~E.\ 2019, \mnras, 484, 3451 
\bibitem[Milisavljevic et al.(2017)]{milisavljevic17} Milisavljevic, D., Patnaude, D.~J., Raymond, J.~C., et al.\ 2017, \apj, 846, 50 
\bibitem[Modjaz et al.(2016)]{modjaz16} Modjaz, M., Liu, Y.~Q., Bianco, F.~B., \& Graur, O.\ 2016, \apj, 832, 108 
\bibitem[Modjaz et al.(2019)]{modjaz19} Modjaz, M., Bianco, F.~B., Siwek, M., et al.\ 2019, arXiv:1901.00872 
\bibitem[Montero-Dorta \& Prada(2009)]{montero09} Montero-Dorta, A.~D., \& Prada, F.\ 2009, \mnras, 399, 1106 
\bibitem[Moriya et al.(2018)]{moriya18} Moriya, T.~J., Nicholl, M., \& Guillochon, J.\ 2018, \apj, 867, 113 
\bibitem[Nadyozhin(1994)]{nadyozhin94} Nadyozhin, D.~K.\ 1994, \apjs, 92, 527 
\bibitem[Nicholl et al.(2013)]{nicholl13} Nicholl, M., Smartt, S.~J., Jerkstrand, A., et al.\ 2013, \nat, 502, 346 
\bibitem[Nicholl et al.(2015)]{nicholl15} Nicholl, M., Smartt, S.~J., Jerkstrand, A., et al.\ 2015, \apjl, 807, L18
\bibitem[Nicholl et al.(2016a)]{nicholl16a} Nicholl, M., Berger, E., Smartt, S.~J., et al.\ 2016, \apj, 826, 39 
\bibitem[Nicholl et al.(2016b)]{nicholl16b} Nicholl, M., Berger, E., Margutti, R., et al.\ 2016, \apjl, 828, L18 
\bibitem[Nicholl et al.(2017)]{nicholl17} Nicholl, M., Guillochon, J., \& Berger, E.\ 2017, \apj, 850, 55 
\bibitem[Nicholl et al.(2018a)]{nicholl18a} Nicholl, M., Berger, E., Blanchard, P.~K, Gomez, S., \& Chornock, R.\ 2018, arXiv:1808.00510 
\bibitem[Nicholl (2018b)]{nicholl18b} Nicholl, M.\ 2018, Research Notes of The AAS, 2, 230
\bibitem[Nicholl et al.(2018c)]{nicholl18c} Nicholl, M., Blanchard, P.~K., Berger, E., et al.\ 2018, \apjl, 866, L24 
\bibitem[Oliphant (2006)]{oliphant06} Oliphant, T. E.\ 2006, Trelgol Publishing
\bibitem[Owocki(2015)]{owocki15} Owocki, S.~P.\ 2015, Very Massive Stars in the Local Universe, 412, 113 
\bibitem[Perets et al.(2010)]{perets10} Perets, H.~B., Gal-Yam, A., Mazzali, P.~A., et al.\ 2010, \nat, 465, 322 
\bibitem[Quimby et al.(2011)]{quimby11} Quimby, R.~M., Kulkarni, S.~R., Kasliwal, M.~M., et al.\ 2011, \nat, 474, 487
\bibitem[Quimby et al.(2018)]{quimby18} Quimby, R.~M., De Cia, A., Gal-Yam, A., et al.\ 2018, \apj, 855, 2 
\bibitem[Rakavy \& Shaviv(1967)]{rakavy67} Rakavy, G., \& Shaviv, G.\ 1967, \apj, 148, 803 
\bibitem[Rezaeikh et al.(2014)]{rezaeikh14} Rezaeikh, S., Javadi, A., Khosroshahi, H., \& van Loon, J.~T.\ 2014, \mnras, 445, 2214 
\bibitem[Schlafly \& Finkbeiner(2011)]{SF2011} Schlafly, E.~F., \& Finkbeiner, D.~P.\ 2011, \apj, 737, 103
\bibitem[Schmidt et al.(1989)]{schmidt89} Schmidt, G.~D., Weymann, R.~J. \& Foltz, C.~B.\ 1989, Publications of the Astronomical Society of the Pacific, 101, 713
\bibitem[Science Software Branch at STScI(2012)]{science12} Science Software Branch at STScI 2012, Astrophysics Source Code Library, ascl:1207.011 
\bibitem[Smith et al.(2016)]{smith16} Smith, M., Sullivan, M., D'Andrea, C.~B., et al.\ 2016, \apjl, 818, L8 
\bibitem[Smithsonian Astrophysical Observatory(2000)]{sao00} Smithsonian Astrophysical Observatory 2000, Astrophysics Source Code Library, ascl:0003.002 
\bibitem[Stevenson et al.(2016)]{stevenson16} Stevenson, K.~B., Bean, J.~L., Seifahrt, A., et al.\ 2016, \apj, 817, 141 
\bibitem[Stritzinger et al.(2009)]{stritzinger09} Stritzinger, M., Mazzali, P., Phillips, M.~M., et al.\ 2009, \apj, 696, 713 
\bibitem[Taddia et al.(2018)]{taddia18} Taddia, F., Sollerman, J., Fremling, C., et al.\ 2018, arXiv:1806.10000 
\bibitem[Taddia et al.(2019)]{taddia19} Taddia, F., Sollerman, J., Fremling, C., et al.\ 2019, \aap, 621, A71 
\bibitem[Valenti et al.(2008)]{valenti08} Valenti, S., Elias-Rosa, N., Taubenberger, S., et al.\ 2008, \apjl, 673, L155 
\bibitem[Vreeswijk et al.(2017)]{vreeswijk17} Vreeswijk, P.~M., Leloudas, G., Gal-Yam, A., et al.\ 2017, \apj, 835, 58 
\bibitem[Woosley et al.(2007)]{woosley07} Woosley, S.~E., Blinnikov, S., \& Heger, A.\ 2007, \nat, 450, 390 
\bibitem[Woosley(2017)]{woosley17} Woosley, S.~E.\ 2017, \apj, 836, 244 
\bibitem[Wyrzykowski et al.(2012)]{wyrzykowski12} Wyrzykowski, {\L}., Hodgkin, S., Blogorodnova, N., Koposov, S., \& Burgon, R.\ 2012, 2nd Gaia Follow-up Network for Solar System Objects, 21 
\bibitem[Yan et al.(2017)]{yan17} Yan, L., Lunnan, R., Perley, D.~A., et al.\ 2017, \apj, 848, 6 
\bibitem[Yoshida et al.(2016)]{yoshida16} Yoshida, T., Umeda, H., Maeda, K., \& Ishii, T.\ 2016, \mnras, 457, 351 
\bibitem[Yusof et al.(2013)]{yusof13} Yusof, N., Hirschi, R., Meynet, G., et al.\ 2013, \mnras, 433, 1114 

\end{thebibliography}
\end{document}